\documentclass[11pt]{article}

\usepackage{epsfig}
\begin{document}
%******************************************************************
%\baselineskip=.33in
%******************************************************************

%***********************
\newcommand{\be}{\begin{equation}}
\newcommand{\ee}{\end{equation}}
\newcommand{\bea}{\begin{eqnarray}}
\newcommand{\eea}{\end{eqnarray}}
\newcommand{\da}{\dagger}
\newcommand{\dg}[1]{\mbox{${#1}^{\dagger}$}}
\newcommand{\hlf}{\mbox{$1\over2$}}
\newcommand{\lfrac}[2]{\mbox{${#1}\over{#2}$}}
\newcommand{\scsz}[1]{\mbox{\scriptsize ${#1}$}}
\newcommand{\tsz}[1]{\mbox{\tiny ${#1}$}}
%***************

\begin{flushright} 
% gr-qc/0411082 
\end{flushright} 

\begin{center}

\Large{\bf Laser Ranging to the Moon, Mars and Beyond}

\vspace{0.4in}

\normalsize
\bigskip 

Slava G. Turyshev, James G. Williams, Michael Shao, John D. Anderson\\

{\it Jet Propulsion Laboratory, California Institute of  Technology, \\
4800 Oak Grove Drive, 
Pasadena, CA 91109, USA}

\vskip 15pt
Kenneth L. Nordtvedt, Jr.  \\

{\it Northwest Analysis, 118 Sourdough Ridge Road, Bozeman, MT 59715 USA}\\
\vskip 15pt

Thomas W. Murphy, Jr.\\
{\it Physics Department, University of California, San Diego \\ 
9500 Gilman Dr., La Jolla, CA 92093 USA}

\normalsize
\vskip 15pt

%***********************************************
% \today
%**************************************************

\vskip 25pt
%\vspace{0.25in}
%\bigskip 

\end{center}

%*******************************************************************
%\baselineskip=.33in
%******************************************************************

\begin{abstract}

Current and future optical technologies will aid exploration of the Moon and Mars while advancing fundamental physics research in the solar system. Technologies and possible improvements in the laser-enabled tests of various physical phenomena are considered along with a space architecture that could be the cornerstone for robotic and human exploration of the solar system. In particular, accurate ranging to the Moon and Mars would not only lead to construction of a new space communication infrastructure enabling an improved navigational accuracy, but will also provide a significant improvement in several tests of gravitational theory: the equivalence principle, geodetic precession, PPN parameters $\beta$ and $\gamma$, and possible variation of the gravitational constant $G$. Other tests would become possible with an optical architecture that would allow proceeding from meter to centimeter to millimeter range accuracies on interplanetary distances.  This paper discusses the current state and the future improvements in the tests of relativistic gravity with Lunar Laser Ranging (LLR).  We also consider precision gravitational tests with the future laser ranging to Mars and discuss optical design of the proposed Laser Astrometric Test of Relativity (LATOR) mission.  We emphasize that already existing capabilities can offer significant improvements not only in the tests of fundamental physics, but may also establish the infrastructure for space exploration in the near future. Looking to future exploration, what characteristics are desired for the next generation of ranging devices, what is the optimal architecture that would benefit both space exploration and fundamental physics, and what fundamental questions can be investigated? We try to answer these questions.

\end{abstract}

\vspace{0.2in}

%\noindent PACS:  04.80.-y, 95.10.Eg, 95.55.Pe

%\newpage

%*********************************
\section{Introduction}

The recent progress in fundamental physics research was enabled by
significant advancements in many technological areas with one of the examples being the continuing development of the NASA Deep Space Network -- critical infrastructure for precision navigation and communication in space. A demonstration of such a progress is the recent Cassini solar conjunction experiment \cite{Bertotti_Iess_Tortora_2003,Anderson_Lau_Giampieri_2004} that was possible only because of the use of Ka-band ($\sim33.4$~GHz) spacecraft radio-tracking capabilities. The  experiment was part of the ancillary science program -- a by-product of this new radio-tracking technology. Becasue of a much higher data rate transmission and, thus, larger data volume delivered from large distances the higher communication frequency was a very important mission capability. The higher frequencies are also less affected by the dispersion in the solar plasma, thus allowing a more extensive coverage, when depp space navigation is concerned. There is still a possibility of moving to even higher radio-frequencies, say to $\sim$60~GHz, however, this would put us closer to the limit that the Earth's atmosphere imposes on signal transmission. Beyond these frequencies radio communication with distant spacecraft will be inefficient.  The next step is switching to optical communication.

Lasers---with their spatial coherence, narrow spectral emission, high power, and well-defined spatial modes---are highly useful for many space applications. While in free-space, optical laser communication (lasercomm) would have an advantage as opposed to the conventional radio-communication methods. Lasercomm would provide not only significantly higher data rates (on the order of a few Gbps), it would also allow a more precise navigation and attitude control. The latter is of great importance for manned missions in accord the ``Moon, Mars and Beyond'' Space Exploration Initiative. In fact, precision navigation, attitude control, landing, resource location, 3-dimensional imaging, surface scanning,  formation flying and  many other areas are thought
only in terms of laser-enabled technologies.  Here we investigate how a near-future free-space optical communication architecture might benefit progress in gravitational and fundamental physics experiments performed in the solar system.

This paper focuses on current and future optical technologies and
methods that will advance fundamental physics research in the context of solar system exploration.  There are many activities that focused on the design on an optical transceiver system which will work at the distance comparable to that between the Earth and Mars, and test it on the Moon. This paper summarizes required capabilities for such a system.   In particular, we discuss how  accurate laser ranging to the neighboring celestial bodies, the Moon and Mars, would not only lead to construction of a new space communication infrastructure with much improved navigational accuracy, it will also provide a significant improvement in several tests of gravitational theory. Looking to future exploration, we address the characteristics that are desired for the next generation of ranging devices; we will focus on optimal architecture that would benefit both space exploration and fundamental physics, and discuss the questions of critical importance that can be investigated.

This paper is organized as follows: Section \ref{sec:llr} discusses
the current state and future performance expected with the LLR
technology. Section \ref{sec:mars} addresses the possibility of improving tests of gravitational theories with laser ranging to Mars. Section \ref{sec:interplanetary} addresses the next logical step---interplanetary laser ranging. We discuss the mission proposal for the Laser Astrometric Test of Relativity (LATOR). We present a design for its optical receiver system. Section \ref{sec:tech} addresses a proposal for new multi-purpose space architecture based on optical communication. We present a preliminary design and discuss implications of this new proposal for tests of fundamental physics.  We close with  a summary and recommendations.

\section{LLR Contribution to Fundamental Physics}
\label{sec:llr}

During more than 35 years of its existence lunar laser ranging has become a critical technique available for precision tests of gravitational theory. The 20th century progress in three seemingly unrelated areas of human exploration -- quantum optics, astronomy, and human space exploration, led to the construction of this unique interplanetary instrument to conduct very precise tests of fundamental physics.

In this section we will discuss the current state in LLR tests of
relativistic gravity and explore what could be possible in the
near future.

\subsection{Motivation for Precision Tests of Gravity}

The nature of gravity is fundamental to our understanding of the
structure and evolution of the universe. This importance motivates
various precision tests of gravity both in laboratories and in space. Most of the experimental underpinning for theoretical gravitation has come from experiments conducted in the solar system.  Einstein's general theory of relativity (GR) began its empirical success in 1915 by explaining the anomalous perihelion precession of Mercury's orbit, using no adjustable theoretical parameters.  Eddington's observations of the gravitational deflection of light during a solar eclipse in 1919 confirmed the doubling of the deflection angles predicted by GR as compared to Newtonian and Equivalence Principle (EP) arguments.  Following these beginnings, the general theory of relativity has been verified at ever-higher accuracy. Thus, microwave ranging to the Viking landers on Mars yielded an accuracy of $\sim$0.2\% from the gravitational time-delay tests of GR \cite{Shapiro_etal_1977,Reasenberg_etal_1979,Shapiro_etal_1988,Shapiro_1990}. Recent spacecraft and planetary microwave radar observations reached an accuracy of $\sim$0.15\% \cite{Anderson_Williams_2001,Anderson_etal_2002}. The astrometric observations of the deflection of quasar positions with respect to the Sun performed with Very-Long Baseline Interferometry (VLBI) improved the accuracy of the tests of gravity to $\sim$0.045\% \cite{Robertson_etal_1991,Shapiro_SS_etal_2004}. Lunar Laser Ranging (LLR),  the continuing legacy of the Apollo program, has provided verification of GR improving an accuracy to $\sim$0.011\% via precision measurements of the lunar orbit \cite{Williams_Newhall_Dickey_1996a,Williams_Newhall_Dickey_1996b,Nordtvedt_1968a,Nordtvedt_1968b,Nordtvedt_1968c,Nordtvedt_1991,Dickey_etal_1994,Nordtvedt_1998,Anderson_Williams_2001,Williams_Turyshev_Boggs_2004}. The recent time-delay experiments with the Cassini spacecraft at a solar conjunction  have tested gravity to a remarkable accuracy of $0.0023\%$ \cite{Bertotti_Iess_Tortora_2003} in measuring deflection of microwaves by solar gravity.

Thus, almost ninety years after general relativity was born, Einstein's theory has survived every test.  This rare longevity and the absence of any adjustable parameters, does not mean that this theory is absolutely correct, but it serves to motivate more sensitive tests searching for its expected violation.  The solar conjunction experiments with the Cassini spacecraft have dramatically improved the accuracy in the solar system tests of GR \cite{Bertotti_Iess_Tortora_2003}.  The reported accuracy of $2.3\times10^{-5}$ in measuring the Eddington parameter $\gamma$, opens a new realm for gravitational tests, especially those motivated by the on-going progress in scalar-tensor theories of gravity.\footnote{An independent determination of $\gamma$ from the Cassini Doppler data is in progress at JPL by one of us (JDA) along with Eunice L. Lau and Giacomo Giampieri \cite{Anderson_Lau_Giampieri_2004}. The data are independent and are drawn from JPL's navigation archives (Block V receivers at DSN station DSS25). The DSN's advanced water vapor radiometry (AWVR) data are common to the two analysis groups, but independently reduced to Doppler shift from atmospheric refraction. The JPL analysis suggests a negative bias to $\gamma$, rather than the positive bias of the published result \cite{Bertotti_Iess_Tortora_2003}. Both determinations are consistent with General Relativity at the one-sigma level, $\sim$ 3 $\times$ 10$^{-5}$.} In particular, scalar-tensor extensions of gravity that are consistent with present cosmological models \cite{Damour_Nordtvedt_1993a,Damour_Nordtvedt_1993b, Damour_Polyakov_1994a,Damour_Polyakov_1994b,Damour_Piazza_Veneziano_2002a,Damour_Piazza_Veneziano_2002b,Nordtvedt_2003} predict deviations of this parameter from its GR value of unity at levels of 10$^{-5}$ to 10$^{-7}$.  Furthermore, the continuing inability to unify gravity with the other forces indicates that GR should be violated at some level.  The Cassini result together with these theoretical predictions motivate new searches for possible GR violations; they also provide a robust theoretical paradigm and constructive guidance for experiments that would push beyond the present experimental accuracy for parameterized post-Newtonian (PPN) parameters (for details on the PPN formalism see \cite{Will_2001}).  Thus, in addition to experiments that probe the GR prediction for the curvature of the gravity field (given by parameter $\gamma$), any experiment pushing the accuracy in measuring the degree of non-linearity of gravity superposition (given by another Eddington parameter $\beta$) will also be of great interest. This is a powerful motive for tests of gravitational physics phenomena at improved accuracies.

Analyses of laser ranges to the Moon have provided increasingly stringent limits on any violation of the Equivalence Principle (EP); they also enabled very accurate measurements for a number of relativistic gravity parameters.

\subsection{LLR History and Scientific Background}

\begin{figure}[!t]
 \begin{center}
\noindent    
\psfig{figure=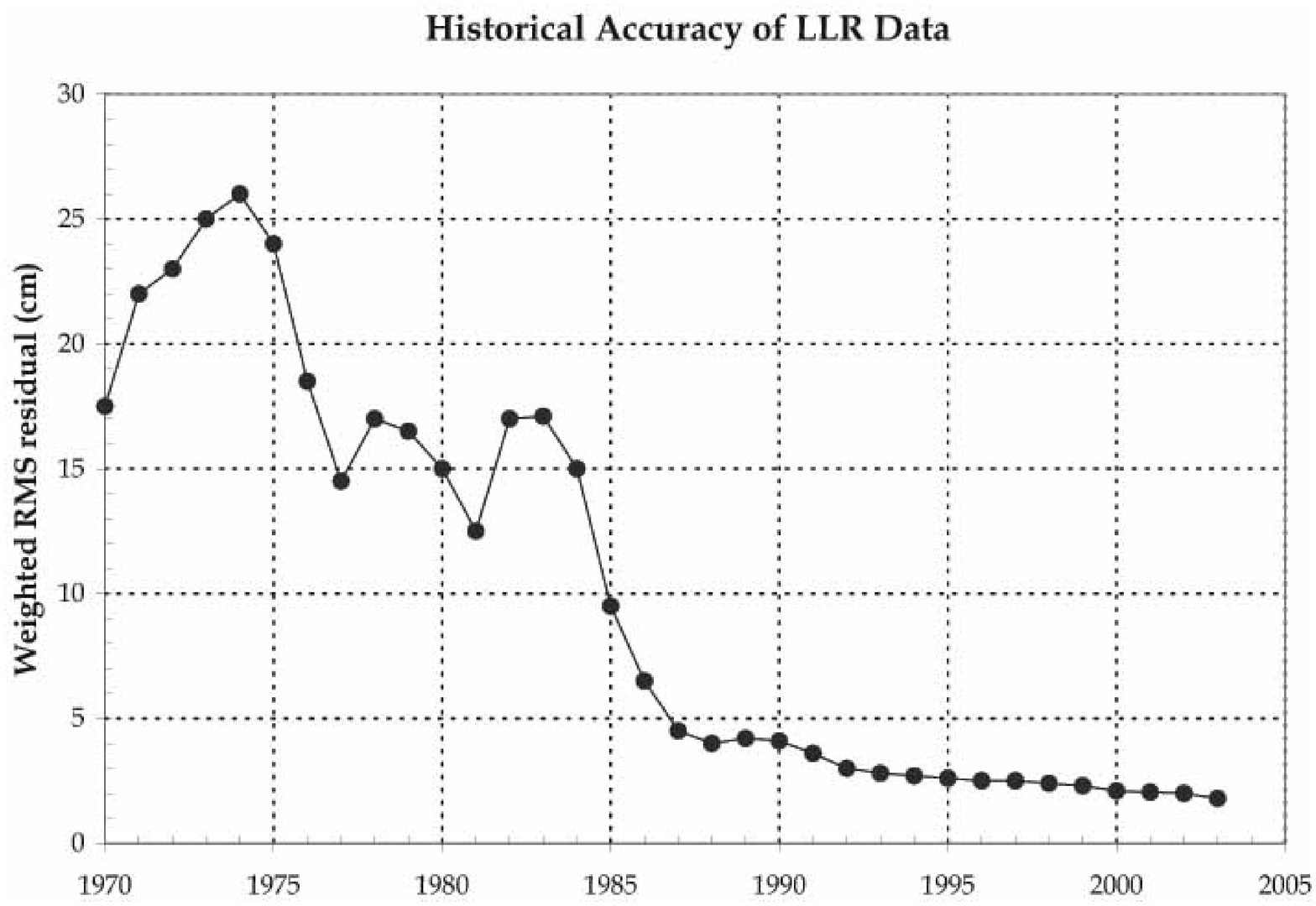,width=145mm}%,height=90mm}
\end{center}
\vskip -10pt 
  \caption{Historical accuracy of LLR data from 1970 to 2004.  
 \label{fig:accuracy}}
\end{figure}

LLR has a distinguished history \cite{Dickey_etal_1994,Bender_etal_1973} dating back to the placement of a retroreflector array on the lunar surface by the Apollo 11 astronauts. Additional reflectors were left by the Apollo 14 and Apollo 15 astronauts, and two French-built reflector arrays were placed on the Moon by the Soviet Luna 17 and Luna 21 missions. Figure \ref{fig:accuracy} shows the weighted RMS residual for each year. Early accuracies using the McDonald Observatory's 2.7 m telescope hovered around 25 cm.  Equipment improvements decreased the ranging uncertainty
to $\sim$15 cm later in the 1970s.  In 1985 the 2.7 m ranging system was replaced with the McDonald Laser Ranging System (MLRS).  In the 1980s ranges were also received from Haleakala Observatory on the island of Maui in the Hawaiian chain and the Observatoire de la Cote d'Azur (OCA) in France.  Haleakala ceased operations in 1990.  A sequence of technical improvements decreased the range uncertainty to the current $\sim$ 2 cm.  The 2.7 m telescope had a greater light gathering capability than the newer smaller aperture systems, but the newer systems fired more frequently and had a much improved range accuracy.  The new systems do not distinguish returning photons against the bright background near full Moon, which the 2.7 m telescope could do, though there are some modern
eclipse observations.

%************

%**************

The lasers currently used in the ranging operate at 10 Hz, with a pulse width of about 200 psec; each pulse contains $\sim10^{18}$ photons. Under favorable observing conditions a single reflected photon is detected once every few seconds. For data processing, the ranges represented by the returned photons are statistically combined into normal points, each normal point comprising up to $\sim100$ photons. There are 15553 normal points are collected until March 2004. The measured round-trip travel times $\Delta t$ are two way, but in this paper equivalent ranges in length units are $c\Delta t/2$. The conversion between time and length (for distance, residuals, and data accuracy) uses 1 nsec=15 cm. The ranges of the
early 1970s had accuracies of approximately 25 cm. By 1976
the accuracies of the ranges had improved to about 15 cm.
Accuracies improved further in the mid-1980s; by 1987 they
were 4 cm, and the present accuracies are $\sim$ 2 cm. One
immediate result of lunar ranging was the great improvement in the accuracy of the lunar ephemeris \cite{Williams_Newhall_Dickey_1996a} and lunar science \cite{cospar_llr_04}. 

LLR measures the range from an observatory on the Earth to a
retroreflector on the Moon.  For the Earth and Moon orbiting the Sun,
the scale of relativistic effects is set by the ratio $(GM/rc^2)\simeq
v^2/c^2\sim10^{-8}$.  The center-to-center distance of the Moon from
the Earth, with mean value 385,000 km, is variable due to such things as eccentricity, the attraction of the Sun, planets, and the Earth's bulge, and relativistic corrections.  In addition to the lunar orbit, the range from an observatory on the Earth to a retroreflector on the Moon depends on the position in space of the ranging observatory and the targeted lunar retroreflector.  Thus, orientation of the rotation axes and the rotation angles of both bodies are important with tidal distortions, plate motion, and relativistic transformations also coming into play.  To extract
the gravitational physics information of interest it is necessary to
accurately model a variety of effects \cite{Williams_Turyshev_Boggs_2004}.

For a general review of LLR see \cite{Dickey_etal_1994}.
A comprehensive paper on tests of gravitational physics is
\cite{Williams_Newhall_Dickey_1996a}.  A recent test of the EP
is in \cite{Anderson_Williams_2001} and other GR tests are in
\cite{Williams_etal_2002}.  An overview of the LLR gravitational
physics tests is given by Nordtvedt \cite{Nordtvedt_1999}.  Reviews
of various tests of relativity, including the contribution by LLR,
are given in \cite{Will_1990,Will_2001}.  Our recent paper describes
the model improvements needed to achieve mm-level accuracy for LLR
\cite{Williams_Turyshev_Murphy_2004}. The most recent LLR results are
given in  \cite{Williams_Turyshev_Boggs_2004}.

\subsection{Tests of Relativistic Gravity with LLR}

LLR offers very accurate laser ranging (weighted rms currently $\sim$ 2 cm or $\sim 5\times 10^{-11}$ in fractional accuracy) to retroreflectors on the Moon. Analysis of these very precise data contributes to many areas
of fundamental and gravitational physics. Thus, these high-precision
studies of the Earth-Moon-Sun system provide the most sensitive tests
of several key properties of weak-field gravity, including Einstein's
Strong Equivalence Principle (SEP) on which general relativity rests (in fact, LLR is the only current test of the SEP). LLR data yielded the strongest limits to date on variability of the gravitational constant (the way gravity is affected by the expansion of the universe), and the best measurement of the de Sitter precession rate.  In this Section we discuss these tests in more details.

\subsubsection{Tests of the Equivalence Principle}
\label{sec:sep_llr}

The Equivalence Principle, the exact correspondence of gravitational and inertial masses, is a central assumption of general relativity and a unique feature of gravitation. EP tests can therefore be viewed in two contexts: tests of the foundations of general relativity, or as searches for new physics.  As emphasized by Damour \cite{Damour_1996,Damour_2001}, almost all extensions to the standard model of particle physics (with best known extension offered by string theory) generically predict new forces that would show up as apparent violations of the EP. 

The weak form the EP (the WEP) states that the gravitational properties of strong and electro-weak interactions obey the EP. In this case the relevant test-body differences are their fractional nuclear-binding differences, their neutron-to-proton ratios, their atomic charges, etc. General relativity, as well as other metric theories of gravity, predict that the WEP is exact. However, extensions of the Standard Model of Particle Physics that contain new macroscopic-range quantum fields predict quantum exchange forces that will generically violate the WEP because they couple to generalized `charges' rather than to mass/energy as does gravity \cite{Damour_Polyakov_1994a,Damour_Polyakov_1994b}. WEP tests can be conducted with laboratory or astronomical bodies, because the relevant differences are in the test-body compositions. Easily the most precise tests of the EP are made by simply comparing the free fall accelerations, $a_1$ and $a_2$, of different test bodies.  For the case when the self-gravity of the test bodies is negligible and for a uniform external gravity field, with the bodies at the same distance from the source of the gravity, the expression for the Equivalence Principle takes the most elegant form:
\begin{equation}
\frac{\Delta a}{a}\equiv \frac{2(a_1-a_2)}{(a_1+a_2)} = 
\left(\frac{M_G}{M_I}\right)_1 - \left(\frac{M_G}{M_I}\right)_2
\end{equation}
where $M_G$  and $M_I$  represent gravitational and inertial masses of each body. The sensitivity of the EP test is determined by the precision of the differential acceleration measurement divided by the degree to which the test bodies differ (e.g. composition).

The strong form of the EP (the SEP) extends the principle to cover the gravitational properties of gravitational energy itself. In other words it is an assumption about the way that gravity begets gravity, i.e. about the non-linear property of gravitation. Although general relativity assumes that the SEP is exact, alternate metric theories of gravity such as those involving scalar fields, and other extensions of gravity theory, typically violate the SEP \cite{Nordtvedt_1968a,Nordtvedt_1968b,Nordtvedt_1968c,Nordtvedt_1991}. For the SEP case, the relevant test body differences are the fractional contributions to their masses by gravitational self-energy. Because of the extreme weakness of gravity, SEP test bodies that differ significantly must have astronomical sizes. Currently the Earth-Moon-Sun system provides the best arena for testing the SEP.

The development of the parameterized post-Newtonian formalism \cite{Nordtvedt_1968b,Will_1971,Will_Nordtvedt_1972}, allows one to describe within the common framework the motion of celestial bodies in external gravitational fields within a wide class of metric theories of gravity. Over the last 35 years, the PPN formalism has become a useful framework for testing the SEP for extended bodies. In that formalism, the ratio of passive gravitational to inertial mass to the first order is given by \cite{Nordtvedt_1968a,Nordtvedt_1968b}:
\begin{equation}
\left[\frac{M_G}{M_I}\right]_{SEP}=1+\eta\Big(\frac{\cal E}{Mc^2}\Big),
\label{eq:sep}
\end{equation}
where $\eta$ is the SEP violation parameter (discussed below), $M$ is the mass of a body and $\cal E$ is its gravitational binding or self-energy:
\begin{equation}
\Big(\frac{\cal E}{Mc^2}\Big)_B = - \frac{G}{2Mc^2} 
\int_{V_B}d^3{\bf x}d^3{\bf y} \frac{\rho_B({\bf x})\rho_B({\bf y})}{|{\bf x} - {\bf y}|}. 
\label{eq:omega}
\end{equation}
The ratio $({\cal E}/Mc^2)$ is proportional to $M$, and typically it is $\sim10^{-25}$ for bodies of laboratory sizes, so experimental accuracy of a part in $10^{13}$ \cite{Adelberger_2001} sheds no light on how gravitational self-energy contributes to the inertial and gravitational masses of bodies.  Numerical evaluation of the integral of expression Eq.~(\ref{eq:omega}) for the Earth and Moon  \cite{Allen_2000,Williams_Newhall_Dickey_1996a} yielded the following values:
\begin{equation}
\Big(\frac{{\cal E}}{Mc^2}\Big)_E=-4.64\times 10^{-10} \hskip 20 pt 
{\rm and} \hskip 20 pt 
\Big(\frac{{\cal E}}{Mc^2}\Big)_m=-0.19\times 10^{-10},
\label{eq:earth_moon}
\end{equation}	
where the subscripts $E$ and $m$ denote the Earth and Moon, respectively.
The relatively small size bodies used in the laboratory experiments possess a negligible amount of gravitational self-energy and therefore such experiments indicate nothing about the equality of gravitational self-energy contributions to the inertial and passive gravitational masses of the bodies \cite{Nordtvedt_1968a}.  
To test the SEP one must utilize planet-sized extended bodies in which case the ratio Eq.~(\ref{eq:omega}) is considerably higher. 

Dynamics of the three-body Sun-Earth-Moon system in the solar system barycentric inertial frame was used to search for the effect of a possible violation of the Equivalence Principle. In this frame, the quasi-Newtonian acceleration of the Moon $(m)$ with respect
to the Earth $(E)$, ${\bf a}={\bf a}_m -  {\bf a}_E$, is calculated to be:
\begin{equation}
{\bf a} = - \mu^* {   {\bf r} \over r^3} + 
 \Bigl ({M_G \over M_I} \Bigl )_{\hskip-2pt m}
\mu_S \Bigl[{{\bf r}_{SE}\over r_{SE}^3} - { {\bf r}_{Sm} \over r_{Sm}^3}\Bigl] + \Bigl[ \Bigl ({M_G \over M_I} \Bigl )_{\hskip-2pt E}-
 \Bigl ({M_G \over M_I} \Bigl )_{\hskip-2pt m}
{\Bigl ]}\mu_S {{\bf r}_{SE} \over r_{SE}^3} \\\label{eq:range1_m}
\end{equation}

\noindent where  $\mu^*\equiv \mu_E (M_G/M_I)_m+ \mu_m (M_G/M_I)_E$ and $\mu_k \equiv G M_k$, also ${\bf r}={\bf r}_m-{\bf r}_E$ and $ {\bf r}_{BC}= {\bf r}_{C}- {\bf r}_{B}$ is the vector from body $B$ to body $C$, $r_{BC}=|{\bf r}_{BC}|$. (The equations that are coded in the LLR integrator are slightly different. For details, consult \cite{Dickey_etal_1994,Williams_Newhall_Dickey_1996a}.) The first term on the hight-hand side of Eq.~(\ref{eq:range1_m}), is the  Newtonian acceleration between the Earth and Moon, ${\bf a}_{N}$, with the second one being the Newtonian tidal acceleration term, ${\bf a}_{tid}$ due to the solar gravity. The last term on the right-hand side is the SEP acceleration term, $ {\bf a}_{\eta}$, which is of order $1/c^2$. While it is not the only term of that order,  the other post-Newtonian $1/c^2$ terms (suppressed in Eq.(\ref{eq:range1_m})) do not affect the determination of $\eta$ until the second post-Newtonian order ($\sim 1/c^4$) \cite{Nordtvedt_1968b,Will_1971,Will_Nordtvedt_1972}. The SEP acceleration is treated as a perturbation on the two-body problem, and the SEP effect is evaluated as an alteration of the planetary Keplerian orbit.

Using expression (\ref{eq:sep}), numerical values (\ref{eq:earth_moon}), to the first order, we obtain from Eq.(\ref{eq:range1_m}),
\begin{equation}
{\bf a} = - \mu^* {   {\bf r} \over r^3} + 
\mu_S \Bigl[{{\bf r}_{SE}\over r_{SE}^3} - { {\bf r}_{Sm} \over r_{Sm}^3}\Bigl ]   +  \eta \Bigl [
\Bigl ( {{\cal E} \over Mc^{2}} \Bigl )_{\hskip-2pt E}- 
\Bigl ({{\cal E} \over Mc^{2}} \Bigl )_{\hskip-2pt m} 
{\Bigl ]}\mu_S {{\bf r}_{SE} \over r_{SE}^3}. \label{eq:range2_m}
\end{equation}

The SEP violation is quantified in the Eqs.~(\ref{eq:sep}) and (\ref{eq:range2_m}) by the parameter $\eta$. In fully-conservative, Lorentz-invariant theories of gravity \cite{Will_1993,Will_2001} the SEP parameter is related to the PPN parameters by 
\begin{equation}
\eta = 4\beta - \gamma - 3.		
\label{eq:eta}
\end{equation}
A difference between gravitational and inertial masses produces potentially observable perturbations in the motion of celestial bodies in the solar system. In general relativity $\eta  = 0$. A unit value for $\eta$ would produce a displacement of the lunar orbit about the Earth, causing a $\sim$13 meter monthly range modulation.  Performing a first-order perturbation analysis of the lunar orbit (assumed circular and planar) in presence of a violation of the equivalence principle, $\eta  \not= 0$  on the dynamics of the Earth-Moon system moving in the gravitational field of the Sun, Nordtvedt \cite{Nordtvedt_1968c} found the first analytical estimate of the corresponding range oscillation \cite{Damour_Vokrouhlicky_1996a,Damour_Vokrouhlicky_1996b}:
\begin{equation}
\delta r =  \eta \Bigl [
\Bigl ( {{\cal E} \over Mc^{2}} \Bigl )_{\hskip-2pt E}- 
\Bigl ({{\cal E} \over Mc^{2}} \Bigl )_{\hskip-2pt m} 
{\Bigl ]} ~\frac{1 + 2n/(n - n')}{n^2 - (n - n')^2}n'^2 ~a' \cos [(n - n') t + D_0]. 
\label{eq:delta_r}
\end{equation}
Here, $n$ denotes the sidereal mean motion of the Moon around the Earth, $n'$ the sidereal mean motion of the Earth around the Sun, and $a'$ denotes the radius of the orbit of the Earth around the Sun (assumed circular). The argument $D=(n - n') t + D_0$ with near synodic period is the mean longitude of the Moon minus the mean longitude of the Sun and is zero at new Moon. (For a more precise derivation of the lunar range perturbation due to the SEP violation acceleration term in Eq.~(\ref{eq:range2_m}) consult \cite{Williams_Newhall_Dickey_1996a}.) Any anomalous radial perturbation will be proportional to $\cos D$. 

Expressed in terms of $\eta$, the radial perturbation in Eq. (\ref{eq:delta_r}) is $\delta r \sim 13 ~\eta~  \cos D$ meters \cite{Nordtvedt_Muller_Soffel_1995,Damour_Vokrouhlicky_1996a,Damour_Vokrouhlicky_1996b}. This effect, generalized to all similar three body situations, the ``SEP-polarization effect.'' LLR investigates the SEP by looking for a displacement of the lunar orbit along the direction to the Sun. The equivalence principle can be split into two parts: the weak equivalence principle tests the sensitivity to composition and the strong equivalence principle checks the dependence on mass.  There are laboratory investigations of the weak equivalence principle (at University of Washington) which are about as accurate as LLR \cite{Baessler_etal_1999,Adelberger_2001}.  LLR is the dominant test of the strong equivalence principle. The most accurate test of the SEP violation effect is presently provided by LLR \cite{Williams_etal_1976,Shapiro_etal_1977,Dickey_etal_1989}, and also in \cite{Dickey_etal_1994,Williams_Newhall_Dickey_1996a,Williams_Newhall_Dickey_1996b,Anderson_Williams_2001}. Recent analysis of LLR data test the EP of $\Delta (M_G/M_I)_{EP} =(-1.0\pm1.4)\times10^{-13}$ \cite{Williams_Turyshev_Boggs_2004}. This result corresponds to a test of the SEP of $\Delta (M_G/M_I)_{SEP} =(-2.0\pm2.0)\times10^{-13}$ with the SEP violation parameter
$\eta=4\beta-\gamma-3$ found to be $\eta=(4.4\pm4.5)\times 10^{-4}$. Using the recent Cassini result for the PPN parameter $\gamma$, PPN parameter $\beta$ is determined at the level of $\beta-1=(1.2\pm1.1)\times 10^{-4}$.

\subsubsection{Other Tests of Gravity with LLR}
\label{sec:llr_other}

LLR data yielded the strongest limits to date on variability of the
gravitational constant (the way gravity is affected by the expansion of the universe), the best measurement of the de Sitter precession rate, and is relied upon to generate accurate astronomical ephemerides. 

The possibility of a time variation of the gravitational constant, {\it G}, was first considered by Dirac in 1938 on the basis of his large number hypothesis, and later developed by Brans and Dicke  in their theory of gravitation (for more details consult \cite{Will_1993,Will_2001}). Variation might be related to the expansion of the Universe, in which case $\dot G/G=\sigma H_0$, where $H_0$ is the Hubble constant, and $\sigma$ is a dimensionless parameter whose value depends on both the gravitational constant and the cosmological model considered. Revival of interest in Brans-Dicke-like theories, with a variable {\it G}, was partially motivated by the appearance of superstring theories where {\it G} is considered to be a dynamical quantity \cite{Marciano1984}. 

Two limits on a change of $G$ come from LLR and planetary ranging.  This is the second most important gravitational physics result that LLR provides. GR does not predict a changing $G$, but some other theories do, thus testing for this effect is important.  The current LLR $\dot G/G=(4\pm9)\times10^{-13}$ yr$^{-1}$ is the most accurate limit published \cite{Williams_Turyshev_Boggs_2004}. The $\dot{G}/G$ uncertainty is 83 times smaller than the inverse age of the universe, $t_0=13.4$ Gyr with the value for Hubble constant $H_0=72$ km/sec/Mpc from the WMAP data \cite{Spergel:2003cb}. The uncertainty for $\dot G/G$ is improving rapidly because its sensitivity depends on the square of the data span.  This fact puts LLR, with its more then 35 years of history, in a clear advantage as opposed to other experiments.

LLR has also provided the only accurate determination of the geodetic
precession.  Ref. \cite{Williams_Turyshev_Boggs_2004} reports a test of geodetic precession, which expressed as a relative deviation from GR, is $K_{gp}=-0.0019\pm0.0064$. The GP-B satellite should provide improved accuracy over this value, if that mission is successfully completed. LLR also has the capability of determining PPN $\beta$ and $\gamma$ directly from the point-mass orbit perturbations.  A future possibility is detection of the solar $J_2$ from LLR data combined with the planetary ranging data.  Also possible are dark matter tests, looking for any departure from the inverse square law of gravity, and checking for a variation of the speed of light.  The accurate LLR data has been able to quickly eliminate several suggested alterations of physical laws.  The precisely measured lunar motion is a reality that any proposed laws of attraction and motion must satisfy.

The above investigations are important to gravitational physics.
The future LLR data will improve the above investigations.  Thus, future LLR data of current accuracy would continue to shrink the uncertainty of $\dot{G}$ because of the quadratic dependence on data span. The equivalence principle results would improve more slowly.  To make a big improvement in the equivalence principle uncertainty requires improved range accuracy, and that is the motivation for constructing the APOLLO ranging facility in New Mexico.

\subsection{Future LLR Data and APOLLO facility}

It is essential that acquisition of the new LLR data will continue in
the future.  Accuracies $\sim$2 cm are now achieved, and further very
useful improvement is expected.  Inclusion of improved data into LLR
analyses would allow a correspondingly more precise determination of
the gravitational physics parameters under study.

LLR has remained a viable experiment with fresh results over 35 years
because the data accuracies have improved by an order of magnitude (see Figure \ref{fig:accuracy}). There are prospects for future LLR station that would provide another order of magnitude improvement. The Apache Point Observatory Lunar Laser-ranging Operation (APOLLO) is a new LLR effort designed to achieve mm range precision and corresponding order-of-magnitude gains in measurements of fundamental physics parameters.  For the first time in the LLR history, using a 3.5 m telescope the APOLLO facility will push LLR into a new regime of multiple photon returns with each pulse, enabling millimeter range precision to be achieved \cite{Murphy_TM_etal_2002, Williams_Turyshev_Murphy_2004}. The anticipated mm-level range accuracy, expected from APOLLO, has a potential to test the EP with a sensitivity approaching 10$^{-14}$. This accuracy would yield sensitivity for parameter $\beta$  at the level of $\sim5\times10^{-5}$ and measurements of the relative change in the gravitational constant, $\dot{G}/G$, would be $\sim$0.1\% the inverse age of the universe.

The overwhelming advantage APOLLO has over current LLR operations is
a 3.5 m astronomical quality telescope at a good site.  The site in
southern New Mexico offers high altitude (2780 m) and very good
atmospheric ``seeing'' and image quality, with a median image resolution of 1.1 arcseconds.  Both the image sharpness and large aperture conspire to deliver more photons onto the lunar retroreflector and receive more of the photons returning from the reflectors, respectively.  Compared to current operations that receive, on average, fewer than 0.01 photons per pulse, APOLLO should be well into the multi-photon regime, with perhaps 5--10 return photons per pulse.  With this signal rate, APOLLO will be efficient at finding and tracking the lunar return, yielding hundreds of times more photons in an observation than current operations deliver. In addition to the significant reduction in statistical error ($\sqrt{N}$
reduction), the high signal rate will allow assessment and elimination
of systematic errors in a way not currently possible.  As a result,
this station is expected to deliver lunar range data accurate to one
millimeter. The APOLLO instrument is expected to acquire its first light ($\sim$ summer 2005) and initiate the regular delivery of LLR data with much improved accuracy.  

\subsection{Emerging Opportunities in the Near Future}

The intent to return to the Moon was announced in January 2004.
NASA is planning to return to the Moon by 2008 with robotic missions
and later with astronauts in the middle of the next decade.  Since LLR
is a valuable working lunar investigation continually operating at the
Moon for more then 35 years, return to the Moon provides an excellent
new opportunity.  What is currently envisioned is a series of missions
to the Moon, including both orbiters and landers.  The primary focus of these planned missions will be lunar exploration and preparation for trips to Mars, but they will also provide opportunities for science, particularly if new reflectors are placed at more widely separated locations (one at zero longitude and zero latitude would also be useful). These new reflectors on the Moon (and later on Mars) can offer significant navigational accuracy for many space vehicles on their approach to the lunar surface or during their flight around the Moon, but they also will contribute significantly to fundamental physics research.

The future of lunar ranging might take two forms, namely passive
retroreflectors and active transponders. The advantages of new
installations of passive retroreflector arrays are their long life and
simplicity. The disadvantages are the weak returned signal and the spread of the reflected pulse arising from lunar librations (apparent changes in orientation of up to 10 degrees).  Insofar as the photon timing error budget is dominated by the libration-induced pulse spread---as is the case in modern lunar ranging---the laser and timing system parameters do not influence the net measurement uncertainty, which simply scales as $1/\sqrt{N}$, where $N$ is the number of photons measured.  Increasing the size of the reflector array by the scale factor $\alpha$ means that the pulse spread also scales as $\alpha$ while the area (and thus $N$) scales as $\alpha^2$.  The net result is no gain or loss in measurement uncertainty.  Corner cubes must be kept individually small to counteract thermal distortions and associated signal loss.  The only way to improve measurement uncertainty in a passive array design is via a sparse array of small corner cubes.  As long as the typical range separation between corner cubes is longer than the laser pulse width and timing uncertainty, each corner cube is individually resolved.  In this way, one can improve data precision via system improvements at the ground station.  A simple approach to the sparse array problem would be to scatter individually mounted corner cubes across a small region of the lunar terrain (laser beam footprint is typically greater than 2 km).  The mounts could be weighted to assure upward-pointing cubes.  Relative positions can be random, and will be sorted out after the fact.

Active laser transponders on the lunar surface are attractive not only because of the strong return and insensitivity to lunar orientation effects, but also because these may become an increasingly important space architecture, as discussed later.  A lunar installation would provide valuable early feedback on their operational characteristics. Active transponders would require power 
and would have more limited lifetimes.  Transponders would probably have internal electronic delays that would need to be calibrated or estimated, and these delays might be temperature sensitive (an unknown temperature variation would correlate with the SEP test.  Transponders can also be used to good effect in asynchronous mode, wherein the received pulse train is not related to the transmitted pulse train, but the transponder unit records the temporal offsets between the two signals.

The LLR retroreflectors are the most accurately known positions on
the Moon. We hope that new retroreflectors and/or transponders will be
delivered to the lunar surface with one of the near future missions. LLR can provide valuable input in determining the optimal location on the Moon for these devices. One can think about the contribution of smaller retroreflector arrays (or random-scatter arrays) for the use on automated spacecraft and larger ones for manned missions. For
transponders, lifetime is important and internal time delays require
thought and good design.  One could also benefit from co-locating
passive and active devices and use a few LLR capable stations ranging
retroreflectors to calibrate the delay vs temperature response of the
transponders (with their more widely observable strong signal).

We shall now discuss the possibility of precision laser ranging to Mars, as the next step in the solar system exploration. 

\section{Laser Ranging to Mars}
\label{sec:mars}

There are three different experiments that can be done with accurate
ranges to Mars: a test of the SEP (similar to LLR), a solar conjunction experiment measuring the deflection of light in the solar gravity, similar to the Cassini experiment, and a search for temporal variation in the gravitational constant {\it G}. The Earth-Mars-Sun-Jupiter system allows for a sensitive test of the SEP which is qualitatively different from that provided by LLR \cite{Anderson_etal_1996}. 
Furthermore, the outcome of these ranging experiments has the potential to improve the values of the two relativistic parameters---a combination of PPN parameters $\eta$ (via test of SEP) and a direct observation of the PPN parameter $\gamma$ (via Shapiro time delay or solar conjunction experiments). (This is quite different compared to LLR, as the small variation of Shapiro time delay prohibits very accurate independent determination of the parameter $\gamma$). The Earth-Mars range would also provide for a very accurate test of $\dot{G}/G$.

This section qualitatively addresses the near-term possibility of laser ranging to Mars and addresses the above three effects.

\subsection{Planetary Test of the SEP with Ranging to Mars}

Earth-Mars ranging data can provide a useful estimate of the SEP parameter $\eta$ given by Eq.~(\ref{eq:eta}).  It was demonstrated in \cite{Anderson_etal_1996} that if future Mars missions provide ranging measurements with an accuracy of $\sigma$ centimeters, after ten years of ranging the expected accuracy for the SEP parameter $\eta$ may be of order $\sigma\times 10^{-6}$. These ranging measurements will also provide the most accurate determination of the mass of Jupiter, independent of the SEP effect test.

It has been observed previously that a measurement of the Sun's
gravitational to inertial mass ratio can be performed using the Sun-Jupiter-Mars or Sun-Jupiter-Earth system \cite{Nordtvedt_1970, Shapiro_Counselman_King_1976,Anderson_etal_1996}. The question we would like to answer here is how accurately can we do the SEP test given the accurate ranging to Mars?  We emphasize that the Sun-Mars-Earth-Jupiter system, though governed basically by the same equations of motion as Sun-Earth-Moon system, is significantly different physically. For a given value of SEP parameter $\eta$ the polarization effects on the Earth and Mars orbits are almost two orders of magnitude larger than on the lunar orbit.  Below we examine the SEP effect on the Earth-Mars range, which has been measured as part of the Mariner 9 and Viking missions with ranging accuracy $\sim 7$ m \cite{Shapiro_etal_1977,Reasenberg_etal_1979,Hellings_etal_1983,Hellings_1984}.  The main motivation for our analysis is the near-future Mars missions that should yield ranging data, accurate to $\sim 1$ cm. This accuracy would bring additional capabilities for the precision tests of fundamental and gravitational physics.

\subsubsection{Analytical Background for a Planetary SEP Test}

The dynamics of the four-body Sun-Mars-Earth-Jupiter system in the Solar system barycentric inertial frame were considered.
The quasi-Newtonian acceleration of the Earth $(E)$ with respect
to the Sun $(S)$, ${\bf a}_{SE}={\bf a}_E -  {\bf a}_S$,  is straightforwardly calculated to be:
\begin{equation}
 {\bf a}_{SE} =  
- \mu^*_{SE}\cdot {   {\bf r}_{SE} \over r_{SE}^3} + 
 \Bigl ({M_G \over M_I} \Bigl )_{\hskip-2pt E}
\sum_{b=M,J} \mu_b   \Bigl [
{  {\bf r}_{bS}\over r_{bS}^3} - { {\bf r}_{bE} \over r_{bE}^3}
\Bigl ]   +  \Bigl [\Bigl ({M_G \over M_I} \Bigl )_{\hskip-2pt S}- 
\Bigl ({M_G \over M_I} \Bigl )_{\hskip-2pt E}
{\Bigl ]} \sum_{b = M,J} \mu_b {  {\bf r}_{bS} \over r_{bS}^3}, \\\label{eq:range1}
\end{equation}

\noindent where  $\mu^*_{SE} \equiv \mu_S (M_G/M_I)_E+ \mu_E (M_G/M_I)_S$ and $\mu_k \equiv G M_k$. The subscripts $(M)$ and $(J)$ indicate Mars and Jupiter, respectively. 

Numerical evaluation of the integral of Eq.~(\ref{eq:omega}) for the standard solar model
\cite{Ulrich_1982} obtains
%\begin{equation}
$({{\cal E} / Mc^2})_S \approx -3.52 \cdot 10^{-6}$, 
%\end{equation}
with $({{\cal E} / Mc^2})_E\approx -4.64 \cdot 10^{-10}$ given by Eq.~(\ref{eq:omega}). These values allow evaluation of the relative strength of the terms in Eq.~(\ref{eq:range1}). The first term on the hight-hand side of Eq.~(\ref{eq:range1}), is the  Newtonian acceleration, ${\bf a}_{N}$, with the second one being the Newtonian tidal acceleration term, ${\bf a}_{tid}$. The last term on the right-hand side is the SEP acceleration term, $ {\bf a}_{\eta}$, which is of order $1/c^2$. While it is not the only term of that order,  the other post-Newtonian $1/c^2$ terms (suppressed in Eq.(\ref{eq:range1})) do not affect the determination of $\eta$ until the second post-Newtonian order ($\sim 1/c^4$). Note, that $a_{\eta}/a_N \sim \eta \cdot10^{-10}$ \hskip2pt and \hskip2pt $a_{tid}/a_N\sim 7 \cdot 10^{-6}$. Given that level of accuracy, we ignore the mutual attraction of the two planets, Earth and Mars. For the purposes of this paper, the SEP acceleration is treated as a perturbation on the restricted three-body problem, and the SEP effect is evaluated as an alteration of the planetary Keplerian orbit.
In addition, noticing that $\mu_M/R_{SM}  \ll \mu_J/R_{SJ}$, we obtain from Eq.(\ref{eq:range1})
\begin{equation}
{\bf a}_{SE} \approx -\mu^*_{SE} \cdot \hskip2pt { {\bf r}_{SE} \over r_{SE}^3} + \mu_J \Bigl [ { {\bf r}_{JS} \over r_{JS}^3}
- {{\bf r}_{JE} \over r_{JE}^3}  \Bigl ] +
  \eta  \Bigl ( {{\cal E} \over Mc^2} \Bigl )_{\hskip-2pt S}
\mu_J {{\bf r}_{JS} \over r_{JS}^3}.
\label{eq:range3}
\end{equation}
\noindent Corresponding equations for Mars are obtained
 by replacing subscript $E$ by $M$ in Eqs.~(\ref{eq:range1}) and (\ref{eq:range3}). To good approximation the SEP violation acceleration ${\bf a}_{\eta}=\eta {\hskip 2pt} ({{\cal E}/ Mc^2})_{\hskip-2pt S}{\hskip 2pt} \mu_J ({{\bf r}_{JS} / r_{JS}^3})$ has constant magnitude of ${a}_{\eta}=\eta {\hskip 2pt} ({{\cal E}/ Mc^2})_{\hskip-2pt S}{\hskip 2pt} \mu_J/ r_{JS}^2 \approx \eta \cdot 7.34 \times 10^{-11}$ cm/s$^2$ and  points in the direction from Jupiter to the Sun, and since it  depends only on the mass distribution in the Sun, the Earth and Mars experience the same perturbing acceleration. The responses of the trajectories of each of these planets  due to the term ${\bf a}_{\eta}$ determine the perturbation in the  Earth-Mars range and allow a detection of the SEP parameter $\eta$ through a ranging experiment.

The presence of the  acceleration term ${\bf a}_\eta$ in the
equations of motion results in a polarization of the orbits of Earth and Mars, exemplifying the planetary SEP effect. We investigate   here the accuracy with which the parameter $\eta$ can be determined through Earth-Mars ranging. 

\subsubsection{Expected Accuracy of a Planetary SEP Test}

Accurate ranging to Mars may be used to search for a violation of the
SEP.  One can determine the PPN parameter $\eta$ from the improved solution for the Martian orbit.  By precisely monitoring the range between the two planets, Earth and Mars, one studies their free-fall accelerations towards the Sun and Jupiter.  The PPN model of this range includes terms due to violation of the SEP introduced by
the possible inequality between gravitational and inertial masses
\cite{Anderson_etal_1996}. Should $\eta$ have a small, but finite value, the Martian orbit will be perturbed by the force responsible for the violation of the SEP.  Assuming circular orbits for the planets (similar to the discussion in Section \ref{sec:sep_llr}), the size of the corresponding Earth-Mars range perturbation is $\delta r \sim 1097 ~\eta$ meters. If one accounts for the orbital eccentricity and inclination effects, together with the tidal interaction, the size of this range perturbation increases by almost a factor of 1.43, reaching magnitude of  $\delta r = 1569 ~\eta$ meters.

If the accuracy in the Earth-Mars range reaches 1 cm, one will be able
to determine the parameter $\eta$ with a single measurement accuracy of $\Delta\eta \sim$ 1 cm/(1569 m) =  $6.4 \times 10^{-6}$.  The term single measurement stands here for the determination of $\eta$ based on range data collected for at least half of the Martian orbit (1.881 years) or $\sim$ 1 Earth year. The accuracy in determining $\eta$ increases, if one is able to continue ranging to Mars with this accuracy for a number of years. For instance, after 10 years (or slightly more than 5 complete Martian years), the experiment will yield $\eta$ with accuracy of $\sim 2 \times 10^{-6}$. This is almost two orders of magnitude improvement compared to the LLR determination. Because of a larger gravitational self-energy of the Sun, $({{\cal E} / Mc^2})_S \approx -3.52 \cdot 10^{-6}$, this accuracy would provide a SEP violation test of $[\Delta(M_G/M_I)]_{SEP} \sim 10^{-13}$, which is comparable to that of the present LLR.  

Obtaining high accuracy measurements of the Martian orbit may, however, present a challenge. The complication comes from the uncertainty on the Martian orbit introduced by the asteroids at the level of a few centimeters. There may be a fundamental limit imposed by the dynamical noise coming from the asteroids, thus, a significant effort may be needed to quantify this limit as well as to develop techniques to overcome this difficulty. One may count on the fact that the orbital motion of Mars also may be used to determine the masses of several asteroids. Ranging measurements between the Earth and a lander on Mars are sensitive to changes in Mars' orbit about the Sun due to the gravitational influence of asteroids. Unless the observation period is sufficiently long, the perturbation from the asteroid cannot be separated from orbital corrections to the mean orbit of Mars. Ranging to the Viking Landers with a mission duration of 6 years and a ranging accuracy of 7 m has allowed reliable mass estimates of three large asteroids with short-period amplitudes larger than the measurement noise \cite{Shapiro_etal_1977,Reasenberg_etal_1979}. Regular range measurements to a lander of 2 m accuracy, combined with the Viking ranging data (a span of almost 30 years), would allow the determination of the masses of several additional asteroids. The asteroid mass estimates can be combined with size estimates from optical, infra-red, radar, and occultation observations to provide density estimates, leading to improved information on their composition.

An additional challenge comes from the uncertainty in the life-time
of the Martian Lander (1-2 years only?). The lasercomm will have
to perform flawlessly for the entire experiment duration (at least
several years), being powered with independent nuclear energy sources,
RTGs. Summing this all this up, we would conservatively put possible
accuracy in determining $\eta$ via precise tast of the SEP with a 1 cm Martian ranges at the level of $\Delta\eta = 5 \times 10^{-6}$. This accuracy in determining the
parameter $\eta$, in combination with parameter $\gamma$ determined
via the time delay experiments to a Lander on Mars (discussed in the
next Section \ref{sec:conjunct}) can provide accuracy on the order of
$\sim 1.25 \times 10^{-6}$ in determining another Eddington parameter,
the parameter $\beta$. More rigorous analysis and covariance studies of this possibility are certainly called for.

\subsection{Solar Conjunction Experiments with Ranging to Mars}
\label{sec:conjunct}

Another relativistic parameter that may be precisely measured with
accurate Earth-Mars ranges is the Eddington's PPN parameter $\gamma$. The measurement may be done during one of solar conjunctions in an experiment similar to that of the Cassini mission in 2002 \cite{Bertotti_Iess_Tortora_2003,Anderson_Lau_Giampieri_2004}. In the conjunction experiments one measures both Shapiro time delay of the signal that is going through the solar gravitational field or the deflection angle to which that signal is deflected by the solar gravity \cite{Shapiro_etal_1977,Reasenberg_etal_1979}. A model that describes the effects of the deflection of light and the light-time delay explicitly depends on the parameter $\gamma$, thus the data analysis
efforts and the solution are reasonably well-understood.

In this Section we discuss a solar conjunction experiment with optical transponder  that could be positioned on one of the future Martian Landers. 

\subsubsection{Analytical Background for a Solar Conjunction Experiment}

From a geometrical point of view the Sun, Earth, and Mars
each curve space in their vicinity to varying degrees. The
effect of this curvature is to increase the round-trip travel
time of a laser pulse. The complete relativistic light-time
expression was derived in heliocentric form by Shapiro  in
1964 \cite{Shapiro_1964} and independently by Holdridge in 1967 \cite{Holdridge_1967}. It was formulated in expanded solar-system barycentric form by Moyer in 1977 \cite{Moyer_1977,Moyer_2003}. 
The portion of Moyer's form due to the Sun and Earth is
{}
\begin{equation}
t_j-t_i =\frac{r_{ij}^B}{c}+{(1+\gamma)} \Big[\frac{\mu_S}{c^3}\ln\Big(\frac{r_i^S+r_j^S+r_{ij}^S+(1+\gamma)\frac{\mu_S}{c^2}}
{r_i^S+r_j^S-r_{ij}^S+(1+\gamma)\frac{\mu_S}{c^2}}\Big)+
\frac{\mu_E}{c^3} \ln\Big(\frac{r_i^E+r_j^E+r_{ij}^E}
{r_i^E+r_j^E-r_{ij}^E}\Big)\Big].
\label{eq:shapiro}
\end{equation}
The first term on the right is the geometric travel time due to
coordinate separation; the remaining two terms represent the
curvature effects due to the Sun and Earth. The complete
equation gives the elapsed coordinate time between two photon
events, where an event is indicated by the subscript $i$ or
$j$. Event $1$ is transmission, event $2$ is reflection, and event $3$
is reception. A latin superscript denotes the origin of a vector:
B is the solar-system barycenter, $S$ is the Sun, and $E$ is
the Earth. In the convention used here, the subscript $i$ represents
the earlier of two photon events, $j$ the later of the two
$(j=i+1)$.  The use of the symbols in the equation is as follows. $r_i^S=|{\bf r}^S_i|$ is the magnitude of the vector from the Sun to photon event $i$ transmission (or reflection) at coordinate time $t_i$; $r_j^S$ has the corresponding meaning for photon event $j$ (reflection or reception). $r_{ij}=|{\bf r}^S_j-{\bf r}^S_i|$ is the magnitude of the difference between the vector from the Sun to photon event $j$ at time $t_j$ and the vector from the Sun to photon event $i$ at time $t_i$. 

When the ray path is near the solar limb, the expression Eq.~(\ref{eq:shapiro}) is greatly simplified.  In particular, the general relativistic correction for range delay in Eq.~(\ref{eq:shapiro}) 
can be approximated to include only the terms due to the solar gravity monopole as
\begin{eqnarray}
\Delta t &=& (1 + \gamma) \frac{2\mu_S}{c^3}\ln\Big(\frac{r_Mr_E}{b^2}\Big),\label{eq:delay_short}
\end{eqnarray}
where $b$ is the impact parameter for the ray path, where $r_M$ is the distance from Sun to Mars, $r_E$ is the distance from Sun to Earth. We will use relation  Eq.~(\ref{eq:delay_short}) for our qualitative discussion below.  A more rigorous analysis would have to be done, including covariance studies in order to explore the potential of a precise ranging to Mars. For a ranging experiment the limiting systematic error, after calibration for plasma, is the trajectory error. Fortunately for a Mars Lander experiment, the orbit of Mars is well known, and thus it is relatively easy to separate the relativistic ranging delay and the trajectory parameters.

Below we will analyze the expected accuracy of such a solar conjunction experiment.

\subsubsection{Expected Accuracy for the Eddington's Parameter $\gamma$}

On the limb of the Sun, the Shapiro delay from a source on the Martian surface is about 250 microseconds (or about 74 km). This effect is inversely proportional to the solar impact parameter. With their S- and X- band communication system, the Viking experiment in 1976 could approach the Sun only to the distance of about 11 solar radii (the solar corona was a limiting factor), which corresponds to a delay of about 156 microseconds (or about 46.8 km).

%************
\begin{figure*}[t!]
 \begin{center}
\noindent    
\psfig{figure=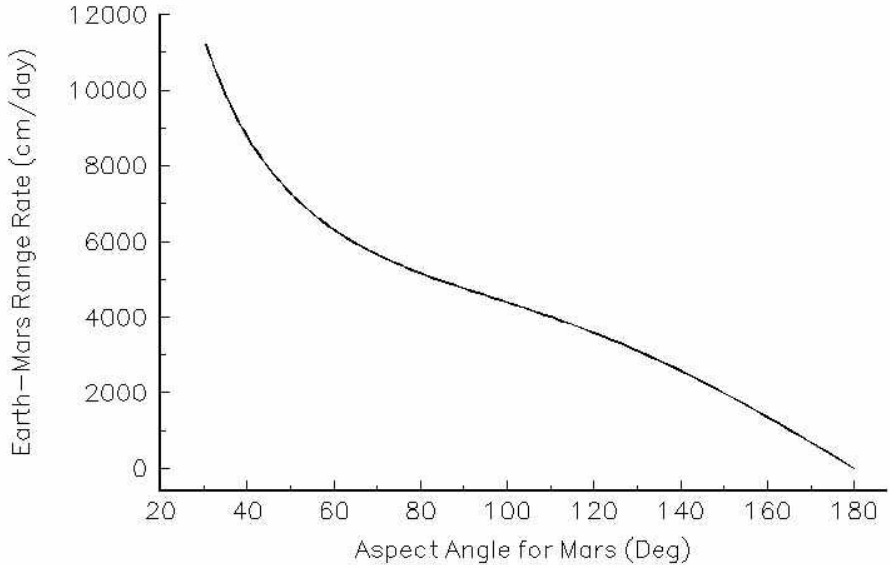,width=120mm}%,height=90mm}
\end{center}
\vskip -14pt 
  \caption{Time rate of change in the Shapiro time delay for a round-trip ranging measurement as a function of aspect angle for Mars. Solar conjunction is at 0$^\circ$, and opposition at 180$^\circ$. Half of this effect comes from $\gamma$, the other half being generically present in any metric theory of gravity.  
 \label{fig:shapiro_delay_mars}}
\end{figure*} 
%**************

With a Cassini-type Ka- and X-band type communication system one can come to the Sun as close as to about 4-6 solar radii, which will result in a delay of 192 microseconds (or about 57.6 km).  (The rate of change in the Shapiro time delay for a round-trip ranging measurement as a function of aspect angle for Mars is given in Figure \ref{fig:shapiro_delay_mars}.) If one measures this delay to 1 cm accuracy, one may determine $\gamma$ accurate to $2 \times 10^{-7}$, which is a bit too optimistic.  On the other hand, the Viking range uncertainty was about 7 meters, which corresponded to $\gamma$ uncertainty on 0.002. If we now measure this range to about 1 cm accuracy, this may lead to improving $\gamma$ as 0.002 $\times$ 1 cm/7 meters =   $3 \times 10^{-6}$, which seem to be reasonable and consistent with possible determination of $\eta$  mentioned above. Also, with a Ka- and X-band ranging system one could come closer to the Sun (the distance may be even closer with an optical link). Therefore, a conservative estimate on the possible determination of $\gamma$ is just slightly below $10^{-6}$, at the level of $8 \times 10^{-7}$. 

The Cassini determination of $\gamma$ came from the range rate data, rather than from only the ranging to the spacecraft, \cite{Anderson_Lau_Giampieri_2004}. In fact, the range accuracy to the Cassini spacecraft was only 7 m, but Doppler velocity determination with the range-rate computation was accurate to 6.25 $\mu$m/s. The range, by itself, adds little to the $\gamma$ determination (the investigator recovered only a Viking result for $\gamma$ of 0.03 with range-only data), but range-rate data does an excellent job. This is because the spacecraft navigational data  was used to describe the time variation of the solar deflection angle as the impact parameter was varied.  

For purposes of modeling a conjunction experiment with Doppler data, one can show that the general relativistic coordinate-independent fractional frequency change $\Delta \nu/\nu$ can be obtained by simply differentiating the coordinate-dependent Eq.~(\ref{eq:shapiro}) with respect to time $t$ to obtain $\Delta v_r = -(1 + \gamma)({2\mu_S}/{c})({db}/{dt})$. The coefficient of $-db/dt$ is just the angular deflection of the ray path as derived from Eq.~(\ref{eq:delay}). When thought of as a series of frequency measurements, a Doppler experiment has more in common with light deflection than it does with Shapiro time delay. In fact, it is a more preferable test of general relativity than the ranging tests, in the sense that the correlation of with the spacecraft trajectory is negligible. 
By carefully monitoring the corresponding Doppler frequency shift and knowing the rate at which the impact parameter was changing, they were able to compute the parameter $\gamma$.  The experiment was not limited by the range accuracy, but rather by the accuracy in determining the velocity of the spacecraft.

Any lasercomm system that will be deployed on the surface of Mars would have to address the issue of the strong background illumination. At the same time, our design studies for the LATOR optical receiver (discussed in Section \ref{sec:lator}) have demonstrated that even a much stronger problem of signal acquisition on the solar background can be solved successfully. Furthermore, the existing design for the LATOR mission would benefit any space optical communication architecture that demands precision navigation and high data rate transmission. 

\subsection{Search for Time Dependence in Gravity Constant with Ranging to Mars}

As pointed out by Dirac, the age of the universe, $H_0^{-1}$, in terms
of the atomic unit of time, $e^2/{mc^3}$, is of order $10^{39}$, as is the ratio, $e^2/(Gm_pm)$, of the electrical force between the electron and proton to the gravitational force between the same particles. This, according to Dirac, suggested that both ratios are
functions of the age of the universe, and that the gravitational constant, $G$, might vary with time. This idea has recently found support in the scalar-tensor theories of gravity which were motivated by the continuing inability to merge gravity with quantum mechanics.  
These theories suggests that the pure tensor gravity of general relativity needs modification or augmentation. The tensor-scalar theories of gravity, where the usual general relativity tensor field coexists with one or several long-range scalar fields, are believed to be the most promising extension of the theoretical foundation of modern gravitational theory. 

The superstring, many-dimensional Kaluza-Klein, and inflationary cosmology theories have revived interest in the so-called `dilaton fields', i.e. neutral scalar fields whose background values determine the strength of the coupling constants in the effective four-dimensional theory. The importance of such theories is that they provide a possible route to the quantization of gravity and unification of physical law. Although the scalar fields naturally appear in the theory, their inclusion predicts different relativistic corrections to Newtonian motions in gravitating systems. In particular, these deviations from general relativity lead to space and time variation of physical ``constants.'' In these theories, the drift in the orbital periods results from time-dependent renormalization of the gravitational constant due to the slowly changing values of the cosmic field (see more details in \cite{Turyshev_etal_2004} and references therein). 

The solar system data was able to provide a very sensitive test to verify this hypothesis.   Monitoring the separation of orbiting bodies offers a possibility to constrain the time variation of $G$. The LLR, planetary and spacecraft ranging data are used to constrain the possible temporal changes because of this effect. To  account for this possibility, and assuming that $\dot{G}$ is constant in first approximation, one writes the effective gravitational constant as 
\begin{equation}
G_{\tt eff}=G[1+\frac{\dot{G}}{G}(t-t_0)],
\label{eq:grav_c}
\end{equation}
 where $G$ and $\dot{G}$ are the values of the gravitational constant and its time derivative at some $t=t_0$. With this $G_{\tt eff}$ in the metric, there appears an additional perturbation in the barycentric equations of motion of planet Mars in the following form:
\begin{equation}
\delta {\bf a}_{\dot{G}}= \frac{\dot{G}}{G} (t-t_0)
\sum_{b\not=M}\mu_a { {\bf r}_{Mb}\over r_{Mb}^3}.
\label{eq:grav_ceq}
\end{equation}
The major contribution to this effect, Eq.~(\ref{eq:grav_ceq}), for the motion of  celestial bodies in the solar system is the Sun. The importance of the highly accurate ranging to Mars  will be in providing the tracking data necessary for establishing a more accurate determination of the  orbital elements of this planet. Then, combined together with other solar system data \cite{Chandler_etal_1994,Anderson_etal_1996,Williams_Turyshev_Boggs_2004}), this information will enable one to perform a more accurate test of the hypothesis (\ref{eq:grav_ceq}).

Similar to LLR data experiment (discussed in Section \ref{sec:llr_other}), analysis of light travel times between Earth and Mars would yields a stringent limit on the fractional variation of the gravitational constant $\dot{G}/G$. The uncertainty for $\dot G/G$ is improving rapidly because its sensitivity depends on the square of the data span.  Continuing of these Earth-Mars laser measurements for five years-even at the accuracy of $\sim 1$ cm would allow for the significant reduction of the uncertainty in $\dot{G}/G$ parameter to 1 part in 10$^{13}$ per year, a limit close to the effect predicted by some theories. This test will be further evaluated in the computer simulations. 

Combining the LLR and Earth-Mars ranging might enable even more accurate test of the gravitational constant.  Both the LLR and the Mars range data have been used to obtain estimates of $\dot{G}/G$.  From the Viking lander, the currently accepted published value of the uncertainty in  this parameter is of order 10$^{-11}$ yr$^{-1}$ \cite{Hellings_etal_1983,Shapiro_1990, Chandler_etal_1994}.  The binary pulsar gives a limit of about $2\times 10^{-11}$ yr$^{-1}$.  The LLR solution in \cite{Williams_Newhall_Dickey_1996a} has an uncertainty of $8 \times 10^{-12}$ yr$^{-1}$ with a value consistent with zero.  This uncertainty is reduced to $1.1 \times 10^{-12}$ yr$^{-1}$ in \cite{Williams_etal_2002} and $9 \times 10^{-12}$ yr$^{-1}$ in \cite{Williams_Turyshev_Boggs_2004}. 
The availability of the accurate ranging data to Mars would enable joint lunar and planetary solutions. It makes sense to combine the planetary and lunar normal equations in joint solutions.  Of particular interest is the possibility to get a better dynamical test of the solar $J_2$ which interferes with planetary gravitational physics tests using Mercury's perihelion precession. A more details analysis of this possibility, including the relevant ranging simulations, will be conducted in the near future. 

\subsection{Expected Performance and An Experiment Concept}

Laser ranging to Mars with a range accuracy $\sim$1 cm can
provide conditions necessary for very precise tests of gravitational
physics. Thus, a relatively short solar conjunction experiment would provide a determination of the Eddington parameter $\gamma$ with
an accuracy of $\sim 8 \times 10^{-7}$, which is already a factor of $\sim$30 improvement compared to the currently available \cite{Bertotti_Iess_Tortora_2003,Anderson_Lau_Giampieri_2004}. For a longer experiment duration, one can also expect to gain a very accurate knowledge of the Martian orbit, which, in return, would help to determine the parameter $\beta$ with an accuracy of $1.25 \times 10^{-6}$. As a result, we will be able to approach the most interesting region where one might see a violation of the general relativity theory due to the presence of an inflationary scalar field. In addition, the expected accuracy in determination of temporal variation of the gravitational constant $\dot{G}/G \sim1\times 10^{-13}$ yr$^{-1}$ is also very interesting for the gravitational and fundamental physics. What is significant is that all three experiments above can be realized with only one lasercomm system deployed at the surface of Mars. This fact motivates a more detailed consideration.  

Our analysis suggest that an experiment that would be able to realize such an excellent opportunity is relatively easy to realize with already existing technology.  The test can be done if a Martian Lander will be equipped with a radio-transponder, independent nuclear power source and a narrow-beam radio antenna to communicate with Earth (an orbiter will have inferior range accuracy). One might think about delivering the lasercomm system to the Martian surface during one of the planned expeditions to Mars. Note that the Earth-Mars distance is prohibitively large for using passive retroreflectors for the lasercomm systems and only active laser transponders are useful for these experiments. Such an excellent accuracy will be, however, compromised by the uncertainty introduced by the asteroids and  solar corona. These dynamical noise factors would affect the accuracy of determination for the Martian orbit. Nevertheless, the expected improvement is still significant and a detailed study of such an experiment had been initiated.

An optical package may use small optics with aperture of 10 cm, a 1 W pulsed laser with a $\sim\mu$rad pointing capabilities. The lasercomm system can be powered with a small Radio-isotope Thermal Generator (RTG) unit for the purposes of independent power supply. (This would be an excellent alternative to the solar panels with a large size, weight and limited lifetime.  However, if power is be provided by solar panels, the experiment can be scaled down to conduct at least one conjunction experiment with a duration of $\sim 25$ days). The entire package can operate for a significant time and can also perform in-situ additional experiments, if needed.  

There may be quite a few opportunities to conduct laser ranging from Mars. Thus,  a one-way laser ranging experiment proposed in the frame of the first Mars Sample Return mission scheduled in 2007 (i.e. ESA's Interplanetary Optical Telemetry project).  One of the additional advantages of such an experiment is its relatively small cost.  Even with a short duration, the results of such an experiment may have a very profound impact on the relativistic gravity tests in the solar system.  The experiment can be conducted at a low cost (all technologies already exist, are proven and space-qualified) and can be launched with one of the next expeditions to Mars, say the Mars 2011 mission. 

The planet Mars has become an object of intensive investigation by many scientists around the world. The next flight opportunity during 2007-09 will continue the series of new space missions to that planet from which we expect to obtain a rich set of data, including spacecraft tracking and planetary radar measurements, and allowing precise relativistic celestial mechanics experiments. Should these mission will accommodate a laser ranging capabilities, there will be potential to significantly improve the tests of gravitational theories with Mars range data accurate to 1 cm.  Anticipating these events, we have estimated the ability for testing SEP violation with Earth-Mars ranging. This analysis shows a rich opportunity for obtaining new scientific results from the program of ranging measurements to Mars.

\section{Interplanetary Laser Ranging}
\label{sec:interplanetary}

The laser ranging may offer a very significant improvements in many areas of deep-space navigation and communication.  What is critical for the purposes of fundamental physics, is that, while in free space, the lasercomm allows for a very precise trajectory estimation and control to the accuracy of less then 1 cm at distances of $\sim 2$ AU. 

In this section we present a concept for the Laser Astrometric Test of Relativity (LATOR) mission that relies on the lasercomm system to perform a very advanced test of general theory relativity. The LATOR mission is a 21st century Michelson-Morley experiment designed to search for presence of a cosmologically-evolved scalar field in the solar system.  This new fundamental physics experiment is designed to test relativistic gravity at an accuracy never achieved before - probing for the first time second-order effects in the gravitational field strength.  As such, LATOR will improve the current accuracy in the tests of general relativity by a factor of 3,000 reaching an accuracy of one part in 10$^8$ (in a single measurement) in measuring space-time curvature around the Sun (as parameterized by the Eddington parameter $\gamma$). 

While primary objective of the LATOR mission is to test GR to the accuracy of at least 3,000 better then currently available, it will also demonstrate the use of the laser ranging technique and optical interferometry for the progress in the interplanetary navigation and communication. 

\subsection{The LATOR Mission}
\label{sec:lator}

When the light deflection in solar gravity is concerned, the magnitude of the first order light deflection effect, as predicted by GR, for the light ray just grazing the limb of the Sun is $\sim1.75$ arcsecond. (Note that 1 arcsecond $\simeq5~\mu$rad; when convenient, below we will use the units of radians and arcseconds interchangeably.) The effect varies inversely with the impact parameter. The second order term is almost six orders of magnitude smaller resulting in  $\sim 3.5$ microarcseconds ($\mu$as) light deflection effect, and which falls off inversely as the square of the light ray's impact parameter \cite{Ken_2PPN_87,lator_cqg2004}. The smallness of the effects emphasize the fact that, among the four forces of nature, gravity is the weakest interaction; it acts at very long distances and controls the large-scale structure of the universe, thus, making the precision tests of gravity a very challenging task. 

The LATOR  mission is designed to directly address the challenges discussed above \cite{lator_cqg2004}. The test will be performed in the solar gravity field using optical interferometry between two micro-spacecraft.  Precise measurements of the angular position of the spacecraft will be made using a fiber coupled multi-chanelled optical interferometer on the ISS with a 100 m baseline. The primary objective of LATOR will be to measure the gravitational deflection of light by the solar gravity to accuracy of 0.1 picoradians (prad) ($\sim0.02 ~\mu$as). 

\begin{figure*}[t!]
 \begin{center}
\noindent    
\psfig{figure=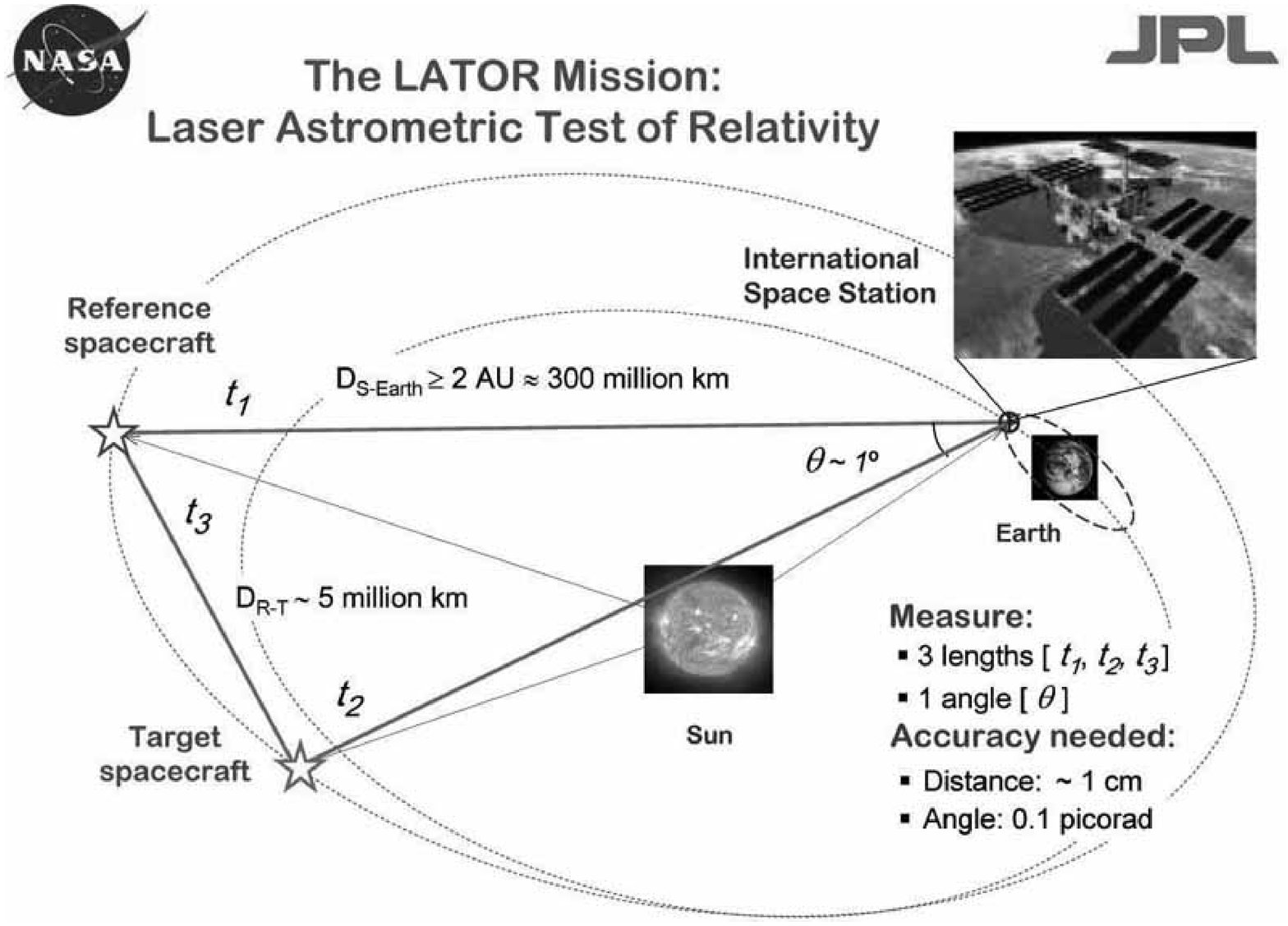,width=158mm}%,height=90mm}
\end{center}
\vskip -10pt 
  \caption{The overall geometry of the LATOR experiment.  
 \label{fig:lator}}
\end{figure*}

The LATOR experiment uses laser interferometry between two micro-spacecraft whose lines of sight pass close by the Sun to accurately measure deflection of light in the solar 
gravity \cite{lator_cqg2004}. Another component of the experimental design is a long-baseline stellar optical interferometer placed on the ISS. Figure \ref{fig:lator} shows the general concept for the LATOR mission including the mission-related geometry, experiment details  and required accuracies. 

%************

%**************

As evident from Figure \ref{fig:lator}, the key element of the LATOR experiment is a redundant geometry optical truss to measure the departure from Euclidean geometry caused by gravity.  Two spacecraft are injected into a heliocentric solar orbit on the opposite side of the Sun from the Earth. The triangle in figure has three independent quantities but three arms are monitored with laser metrology. In particular, each spacecraft equipped with a laser ranging system that enable a measurement of the arms of the triangle formed by the two spacecraft and the ISS.   According to Euclidean rules this determines a specific angle at the interferometer; LATOR can  directly measure this angle directly. In particular, the laser beams transmitted by each spacecraft are detected by a long baseline ($\sim$ 100 m) optical interferometer on the ISS. The actual angle measured at interferometer is compared to angle calculated using Euclidean rules and three side measurements; the difference is the non-Euclidean deflection signal [which varies in time during spacecraft passages] which contains the scientific information.
Therefore, the uniqueness of this mission comes with its built-in geometrically redundant architecture that enables LATOR to measure the departure from Euclidean geometry caused by the solar gravity field to a very high accuracy. The accurate measurement of this departure constitutes the primary mission objective.

In conjunction with laser ranging among the spacecraft and the ISS, LATOR will allow measurements of the gravitational deflection by a factor of more than 3,000 better than had recently been accomplished with the Cassini spacecraft. In particular, this mission will not only measure the key Eddington parameter $\gamma$ to unprecedented levels of accuracy of one part in 10$^8$. The Eddington parameter $\gamma$, whose value in GR is unity, is perhaps the most fundamental PPN parameter, in that $(1-\gamma)$ is a measure, for example, of the fractional strength of the scalar gravity interaction in scalar-tensor theories of gravity \cite{Damour_Nordtvedt_1993a,Damour_Nordtvedt_1993b,Damour_EFarese96,lator_cqg2004}.  Within perturbation theory for such theories, all other PPN parameters to all relativistic orders collapse to their GR values in proportion to $(1-\gamma)$. This is why measurement of the first order light deflection effect at the level of accuracy comparable with the second-order contribution would provide the crucial information separating alternative scalar-tensor theories of gravity from GR \cite{Ken_2PPN_87} and also to probe possible ways for gravity quantization and to test modern theories of cosmological evolution discussed in the previous section.  LATOR is designed to directly address this issue with an unprecedented accuracy; it will also reach ability to measure the next post-Newtonian order ($\propto G^2$) of light deflection with accuracy to 1 part in $10^3$.

LATOR will lead to very robust advances in the tests of fundamental physics: it could discover a violation or extension of GR, or reveal the presence of an additional long range interaction in the physical law.  There are no analogs to the LATOR experiment; it is unique and is a natural culmination of solar system gravity experiments. 

\subsection{The LATOR Mission Design}

To enable the primary objective, LATOR will place two spacecraft into a heliocentric orbit so that observations may be made when the spacecraft are behind the Sun as viewed from the ISS.   The two spacecraft are to be separated by about 1$^\circ$, as viewed from the ISS \cite{lator_cqg2004}. With the help of the JPL Advanced Project Design Team (Team X), we recently conducted a detailed mission design studies. In particular, we analyzed various trajectory options for the deep-space flight segment of LATOR, using both Orbit Determination Program (ODP) and Satellite Orbit Analysis Program (SOAP) --  the two standard JPL navigation software packages. 

The LATOR mission uses laser interferometry between two laser sources placed on separate small spacecraft, whose lines of sight pass close by the Sun, to measure accurately the deflection of light in the solar gravity field.   The two LATOR spacecraft will be placed in a 3:2 Earth resonant orbit that provides three observing sessions during the initial 21 months after the launch with the first session starting in 15 months.  The spacecraft will use the standard Spectrum Astro SA200S bus and will be launched on a Delta II launch vehicle. The nominal mission life time is 22 months with a minimal life of 16 months.

The goal of measuring deflection of light in solar gravity with accuracy of one part in $10^{8}$ requires serious consideration of systematic errors. This work requires a significant effort to properly identify the entire set of factors that may influence the accuracy at this level. Fortunately, we initiated this process being aided with experience of successful development of a number of instruments that require similar technology and a comparable level of accuracy, notably SIM, TPF, Keck and Palomar Testbed Interferometers. This experience comes with understanding various constituents of the error budget, expertize in developing appropriate instrument models; it is also supported by the extensive verification of the expected  performance with the set of instrumental test-beds and existing flight hardware. Details of the LATOR error budget are still being developed and will be published elsewhere, when fully analyzed. 

Here we discuss a the LATOR astrometric observable as it relates to the realization of the future optical infrastructure.

\subsection{Observational Model for LATOR} 

In development of the mission's error budget we use a simple model to capture all error sources and their individual impact on the mission performance \cite{Turyshev_etal_2004}. 
The light paths, $\ell_{ij}$, between the three vortices of the triangle  may be given by an expression for the Shapiro time delay relation, that to the first order in gravitational constant, has the form: 
\begin{equation}
\ell_{ij}=r_{ij}+(1+\gamma)\mu_\odot\ln[\frac{r_i+r_j+r_{ij}}{r_i+r_j-r_{ij}}], ~~~~ \mathbf{r}_{ij}=\mathbf{r}_j-\mathbf{r}_i,
\label{eq:path}
\end{equation}
\noindent where $\mathbf{r}_i$ is the barycentric Euclidian position to one of the three vortices, $i,j\in\{1,3\}$ ($i=3$ is for the ISS), with $r_i=|\mathbf{r}_i|$, being its distance, and $\mu_\odot=GM/c^2$ is the solar gravitational radius. To a similar accuracy, the interferometric delay, $ d_{j}$, for a laser source $j$ has the following approximate form (i.e. differenced Shapiro time delay for the two telescopes separated by an interferometric baseline, $\mathbf{b}$, or $d_j=\ell_{j3}(\mathbf{r}_3)-\ell_{j3}(\mathbf{r}_3+\mathbf{b})$):
{}
\begin{equation}
d_j\simeq(\mathbf{b}\cdot\mathbf{n}_{j3})-(1+\gamma)\mu_\odot 
\frac{2r_jr_3}{r_3+r_j}\frac{\mathbf{b}\cdot(\mathbf{n}_3-\mathbf{n}_{j3})}{p_j^2},
\label{eq:delay}
\end{equation} 
\noindent where $p_j$ is the solar impact parameter for source $j$. Both expressions Eqs.(\ref{eq:path}) and (\ref{eq:delay})  require some additional transformations to keep only the terms with a similar order. The entire LATOR model will account for a whole range of other effects, including terms due to gravitational multipoles, second order deflection, angular momentum contribution, and etc. This work had being initiated and the corresponding results will be reported elsewhere. Below we shall comment only on the conceptual formulation of the LATOR observables. 

The range observations Eq.(\ref{eq:path}) may be used to measure any angle between the three fiducials in the triangle. However, for observations in the solar gravity field, measuring the lengths do not give you a complete information to determine the angles, and some extra information is needed. This information is the mass of the Sun, and, at least one of the impact parameters. Nevertheless, noting that the paths $\underline{\ell}_{ij}$ correspond to the sides of the connected, but gravitationally distorted triangle, one can write $\underline{\ell}_{12}+\underline{\ell}_{23}+\underline{\ell}_{31}=0$, where $\underline{\ell}_{ij}$ is the null geodesic path for light moving between the two points $i$ and $j$. This leads to the expression for the angle between the spacecraft $\cos(\widehat{\underline{\ell}_{31}\underline{\ell}_{32}})=\cos\delta_r =(\ell_{32}^2+\ell_{31}^2-\ell_{12}^2)/(2 \ell_{32} \ell_{31})$. Expression for $\cos\delta_r$ will have both Euclidian and gravitational contributions; their detailed form will not significantly contribute to the discussion below and, thus, it is outside the scope of this paper.

The astrometric observations Eq.(\ref{eq:delay}) will be used to obtain another measurement of the same angle between the two spacecraft. The LATOR interferometer will perform differential observations between the two sources of laser light, measuring the differential delay $\Delta d_{12}=d_2-d_1$ to the accuracy of less than $5$ pm (see below). For the appropriate choice of the baseline orientation, one can present the angle between the two sources of laser light as  $\cos(\widehat{\underline{\ell}_{31}\underline{\ell}_{32}})=\cos\delta_a=1-\Delta d_{12}^2/2b^2$. This expression would have both Euclidian and gravity contributions which are not discussed in detail in this paper.

The two sets of observations obtained by laser ranging and astrometric interferometry form the complete set of LATOR observables. Conceptually,  the LATOR astrometric measurement $\delta_d$ of the gravitational deflection of light may be modeled as
\begin{equation}
\delta_d = \delta_r-\delta_a = c_1\Big(\frac{1}{p} - \frac{1}{p+\Delta p}\Big)+c_2\Big(\frac{1}{p^2} - \frac{1}{(p+\Delta p)^2}\Big),
\end{equation}
 
\noindent where  $\delta_r$ is the angle computed from the range information,  $\delta_a$ is the angle measured astrometrically by the interferometer. $p$ is the impact parameter of the spacecraft closer to the Sun and $\Delta p$ is the difference between the two impact parameters. $c_1$ and $c_2$ are the first and second order terms in the gravitational deflection and are the quantities of interest. Three such measurements are made to simultaneously solve for these constants together with the impact parameter. 
The temporal evolution of the entire triangle structure will produce another set of observables -- the range rate data, expressed as $\delta_v=d\delta_d/dt=(\partial \delta_d/\partial p)dp/dt$ which will also be used to process the data. A fully relativistic model for this additional independent observable, including the contributions of range and angle rates, is being currently developed \cite{Turyshev_etal_2004}.  The error budget is subdivided into three components -- range and interferometer measurements, and spacecraft stability that all relate to the expected performance of the optical system. 

As evident from Figure \ref{fig:lator}, the key element of the LATOR experiment is a redundant geometry optical truss to measure the effects of gravity on the laser signal trajectories.  LATOR will generate four time series of measurements -- one for the optical range of each side of the triangle, plus the angle between light signals arriving at one vertex of the light triangle.  Within the context of a moving Euclidean light triangle, these measurements are redundant.  From a combination of these four times series of data, the several effects of gravity on the light propagations can be precisely and separately determined.  For example: the first and second order gravity monopole deflections go as $1/p$ and $1/p^2$ while the solar quadrupole deflection goes as $1/p^3$ [with $p(t)$ being a laser signal's evolving impact parameter]; the quadrupole moment's deflection has further latitude dependence if spacecraft lines of sight are so located.  The data will be taken over periods in which the laser light's impact parameters $p(t)$ vary from one to ten solar radii, producing time signatures in the data which permits both the separation of the several gravitational effects and the determination of key spacecraft location coordinates needed to calibrate the deflection signals.  The temporal evolution of the entire triangle structure will produce the range rate and angle rate data that will be used to process the experimental data. 

We shall now consider the basic elements of the LATOR optical design. 

\subsection{Optical Design}

\begin{figure*}[t!]
 \begin{center}
\noindent    
\psfig{figure=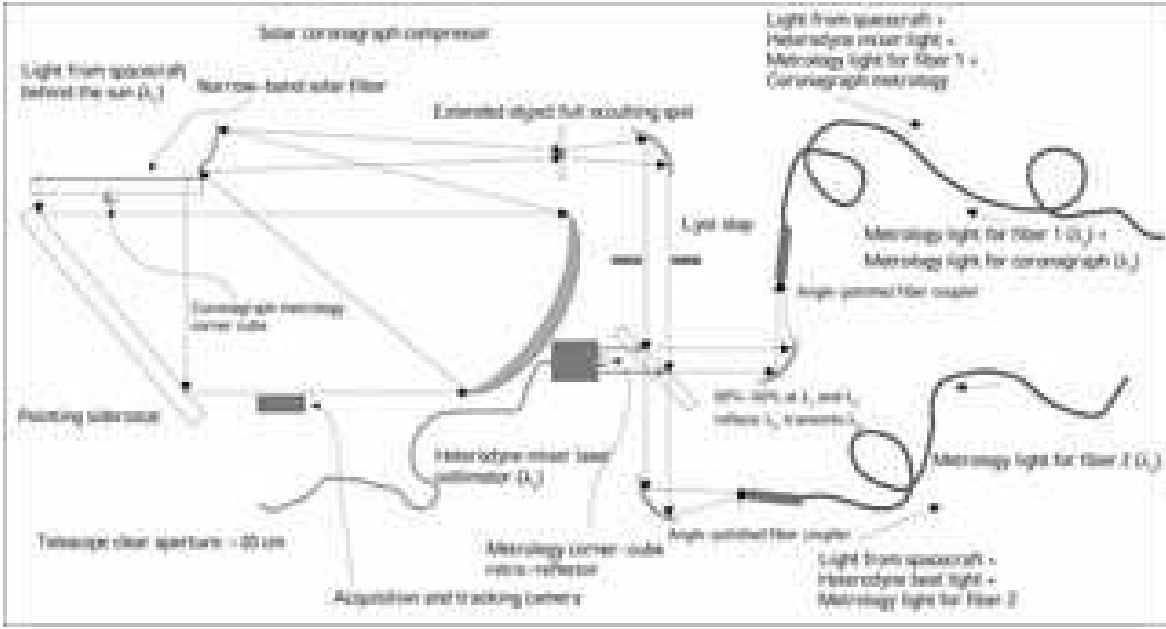,width=155mm}%,height=90mm}
\end{center}
\vskip -10pt 
  \caption{Basic elements of optical design for the LATOR interferometer: The laser light (together with the solar background) is going through a full aperture ($\sim 20$cm) narrow band-pass filter with $\sim 10^{-4}$ suppression properties. The remaining light illuminates the baseline metrology corner cube and falls onto a steering flat mirror where it is reflected to an off-axis telescope with no central obscuration (needed for metrology). It is then enters the solar coronograph compressor by first going through a 1/2 plane focal plane occulter and then coming to a Lyot stop. At the Lyot stop, the background solar light is reduced by a factor of $10^{6}$. The combination of a narrow band-pass filter and coronograph enables the solar luminosity reduction from $V=-26$ to $V=4$ (as measured at the ISS), thus, enabling the LATOR precision observations.
\label{fig:optical_design}}
\end{figure*}

A single aperture of the interferometer on the ISS consists of three 20 cm diameter telescopes (see Figure \ref{fig:optical_design} for a conceptual design). One of the telescopes with a very narrow bandwidth laser line filter in front and with an InGAs camera at its focal plane, sensitive to the 1064 nm laser light, serves as the acquisition telescope to locate the spacecraft near the Sun.

The second telescope emits the directing beacon to the spacecraft. Both spacecraft are served out of one telescope by a pair of piezo controlled mirrors placed on the focal plane. The properly collimated laser light ($\sim$10W) is injected into the telescope focal plane and deflected in the right direction by the piezo-actuated mirrors. 

The third telescope is the laser light tracking interferometer input aperture which can track both spacecraft at the same time. To eliminate beam walk on the critical elements of this telescope, two piezo-electric X-Y-Z stages are used to move two single-mode fiber tips on a spherical surface while maintaining focus and beam position on the fibers and other optics. Dithering at a few Hz is used to make the alignment to the fibers and the subsequent tracking of the two spacecraft completely automatic. The interferometric tracking telescopes are coupled together by a network of single-mode fibers whose relative length changes are measured internally by a heterodyne metrology system to an accuracy of less than 10 pm.

%************
%**************

The spacecraft  are identical in construction and contain a relatively high powered (1 W), stable (2 MHz per hour $\sim$  500 Hz per second), small cavity fiber-amplified laser at 1064 nm. Three quarters of the power of this laser is pointed to the Earth through a 10 cm aperture telescope and its phase is tracked by the interferometer. With the available power and the beam divergence, there are enough photons to track the slowly drifting phase of the laser light. The remaining part of the laser power is diverted to another telescope, which points towards the other spacecraft. In addition to the two transmitting telescopes, each spacecraft has two receiving telescopes.  The receiving telescope, which points towards the area near the Sun, has laser line filters and a simple knife-edge coronagraph to suppress the Sun light to 1 part in $10^4$ of the light level of the light received from the space station. The receiving telescope that points to the other spacecraft is free of the Sun light filter and the coronagraph.

In addition to the four telescopes they carry, the spacecraft also carry a tiny (2.5 cm) telescope with a CCD camera. This telescope is used to initially point the spacecraft directly towards the Sun so that their signal may be seen at the space station. One more of these small telescopes may also be installed at right angles to the first one to determine the spacecraft attitude using known, bright stars. The receiving telescope looking towards the other spacecraft may be used for this purpose part of the time, reducing hardware complexity. Star trackers with this construction have been demonstrated many years ago and they are readily available. A small RF transponder with an omni-directional antenna is also included in the instrument package to track the spacecraft while they are on their way to assume the orbital position needed for the experiment. 

The LATOR experiment has a number of advantages over techniques which use radio waves to measure gravitational light deflection. Advances in optical communications technology, allow low bandwidth telecommunications with the LATOR spacecraft without having to deploy high gain radio antennae needed to communicate through the solar corona. The use of the monochromatic light enables the observation of the spacecraft almost at the limb of the Sun, as seen from the ISS. The use of narrowband filters, coronagraph optics and heterodyne detection will suppress background light to a level where the solar background is no longer the dominant noise source. In addition, the short wavelength allows much more efficient links with smaller apertures, thereby eliminating the need for a deployable antenna. Finally, the use of the ISS will allow conducting the test above the Earth's atmosphere -- the major source of astrometric noise for any ground based interferometer. This fact justifies LATOR as a space mission.

Below we will present the details of the LATOR optical receiver system. 

\subsection{The LATOR Optical Receiver System}
In this Section we present elements for the LATOR optical receiver system.  While we focus on the optics for the two spacecraft, the interferometer has essentially similar optical architecture. 

The LATOR 100 mm receiver optical system is a part of a proposed experiment. This system is located at each of two separate spacecraft placed on heliocentric orbits, as shown in Figure \ref{fig:lator}. The receiver optical system receives optical communication signals form a transmitter on the ISS, that is in orbit around the Earth. To support the primary mission objective, this system must be able to receive the optical communication signal from the uplink system at the ISS that passes through the solar corona at the immediate proximity of the solar limb (at the distance no more then 5 Airy disks). 

%====================================

\begin{table*}[t!]
\begin{center}
\caption{Summary of design parameters for the LATOR optical receiver system.
\label{table:requirements}} \vskip 8pt
\begin{tabular}{rl} \hline\hline
Parameters/Requirements   & Value/Description \\\hline
 & \\[-10pt]
Aperture &  100 mm, unobstructed \\[3pt]
Wavelength & 1064 nm \\[3pt]
Narrow bandpass Filter & 2 nm FWHM over full aperture \\[3pt] 
Focal Planes & APD Data \& CCD Acquisition/Tracking \\[3pt]
APD Field of View & Airy disk field stop (pinhole) in front of APD\\[3pt]
APD Field Stop (pinhole) & Approximately 0.009 mm in diameter \\[3pt] 
APD Detector Size & TBD (a little larger than 0.009 mm)\\[3pt] 
CCD Field of View & 5 arc minutes \\[3pt] 
CCD Detector Size & 640 $\times$ 480 pixels (9.6 mm $\times$ 7.2 mm)\\[3pt]
CCD Detector Pixel Size & 15 $\mu$m\\[3pt] 
Beamsplitter Ratio (APD/CCD) & 90/10\\[3pt] 
Field Stop & `D'-shaped at primary mirror focus \\[3pt] 
Lyot Stop & Circular aperture located at telescope exit pupil\\[3pt] 
\hline\hline
\end{tabular} 
\end{center} 
\end{table*}
%==========================================================  

%************
\begin{figure*}[t!]
 \begin{center}
\noindent    
\psfig{figure=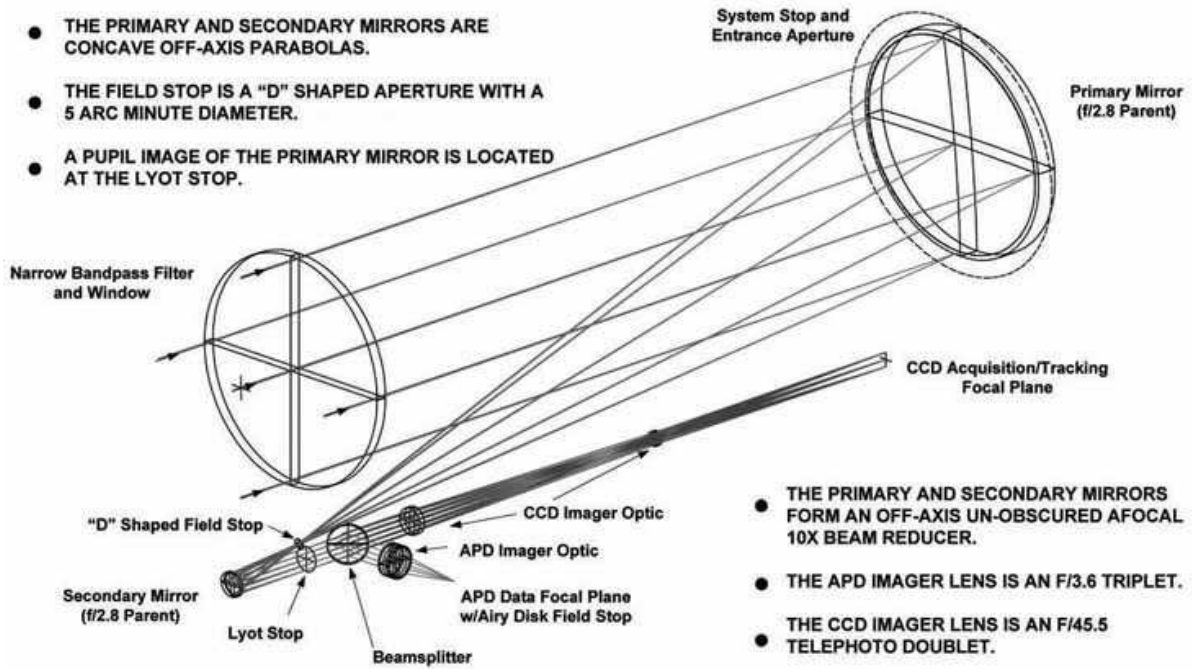,width=158mm}%,height=90mm}
\end{center}
\vskip -10pt 
  \caption{LATOR receiver optical system layout.  
 \label{fig:lator_receiver}}
\end{figure*} 
%**************

Our recent analysis of the LATOR 100 mm receiver optical system successfully satisfied all the configuration and performance requirements (shown in Table \ref{table:requirements}). We have also performed a conceptual design (see Figure \ref{fig:lator_receiver}), which was validated with a CODEV  ray-trace analysis. The ray-trace performance of the designed instrument is diffraction limited in both the APD and CCD channels over the specified field of view at 1064 nm. The design incorporated the required field stop and Layot stop. A preliminary baffle design has been developed for controlling the stray light. 

The optical housing is estimated to have very accommodating dimensions; it measures (500 mm $\times$ 150 mm $\times$ 250 mm). The housing could be made even shorter by reducing the focal length of the primary and secondary mirrors, which may impose some fabrication difficulties. These design opportunities are being currently investigated. 

\subsubsection{Preliminary Baffle Design}

%************
\begin{figure*}[t!]
 \begin{center}
\noindent    
\psfig{figure=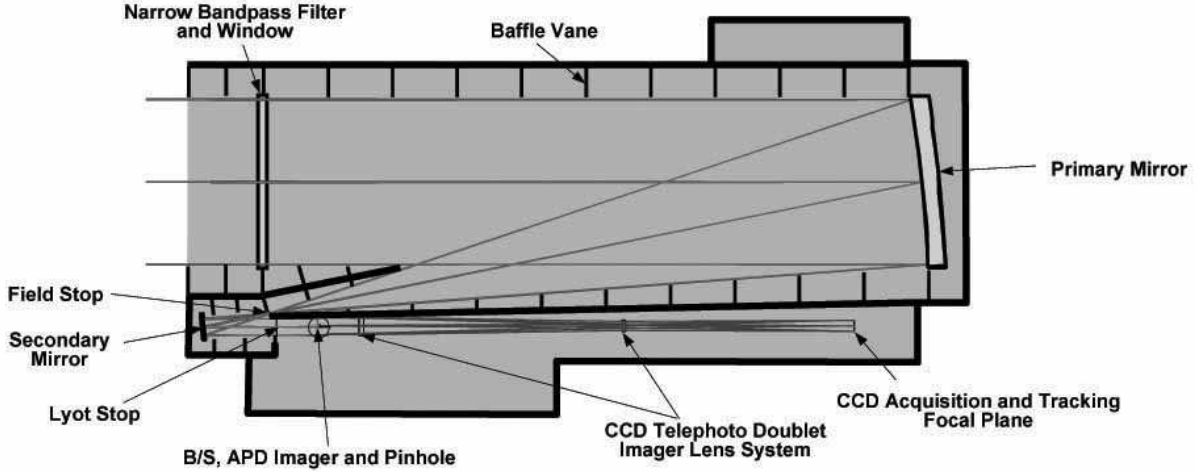,width=160mm}%,height=90mm}
\end{center}
\vskip -10pt 
  \caption{The LATOR preliminary baffle design.  
 \label{fig:lator_buffle}}
\end{figure*} 
%**************

Figure \ref{fig:lator_buffle} shows the LATOR preliminary baffle design. The out-of-field solar radiation falls on the narrow band pass filter and primary mirror; the scattering from these optical surfaces puts some solar radiation into the FOV of the two focal planes. This imposes some requirements on the instrument design.  Thus, the narrow band pass filter and primary mirror optical surfaces must be optically smooth to minimize narrow angle scattering. This may be difficult for the relatively steep parabolic aspheric primary mirror surface. However, the field stop will eliminate direct out-of-field solar radiation at the two focal planes, but it will not eliminate narrow angle scattering for the filter and primary mirror.  Finally, the Lyot stop will eliminate out-of-field diffracted solar radiation at the two focal planes. Additional baffle vanes may be needed several places in the optical system. This design will be further investigated in series of trade-off studies with support from this proposal by also focusing on the issue of stray light analysis. 

\subsubsection{Focal Plane Mapping}

\begin{figure*}[t!]
 \begin{center}
\noindent    
\psfig{figure=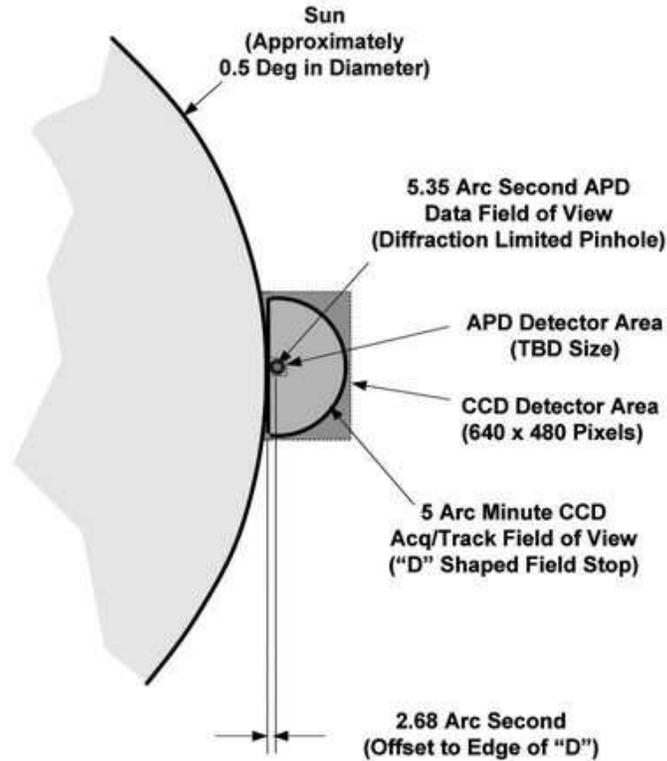,width=92mm}%,height=90mm}
\end{center}
\vskip -10pt 
  \caption{LATOR focal plane mapping (the diagram not to scale).  
 \label{fig:lator_focal}}
\end{figure*}

\begin{figure*}[t!]
 \begin{center}
\noindent    
\psfig{figure=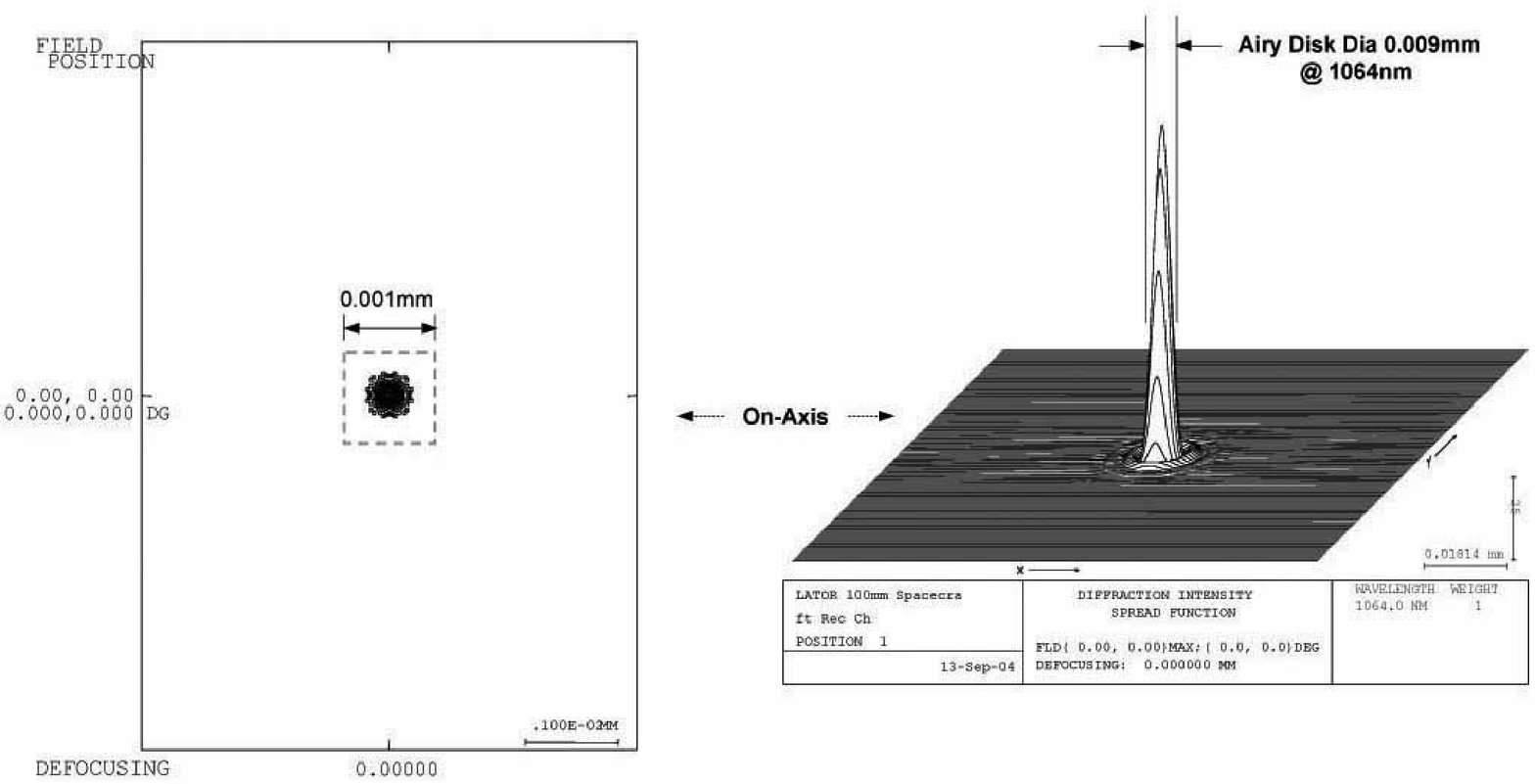,width=158mm}%,height=90mm}
\end{center}
\vskip -10pt 
  \caption{APD channel geometric (left) and diffraction (right) PSF.  
 \label{fig:lator_apd}}
%\end{figure*} 
%%**************
%%************
%\begin{figure*}[t!]
 \begin{center}
\noindent    
\psfig{figure=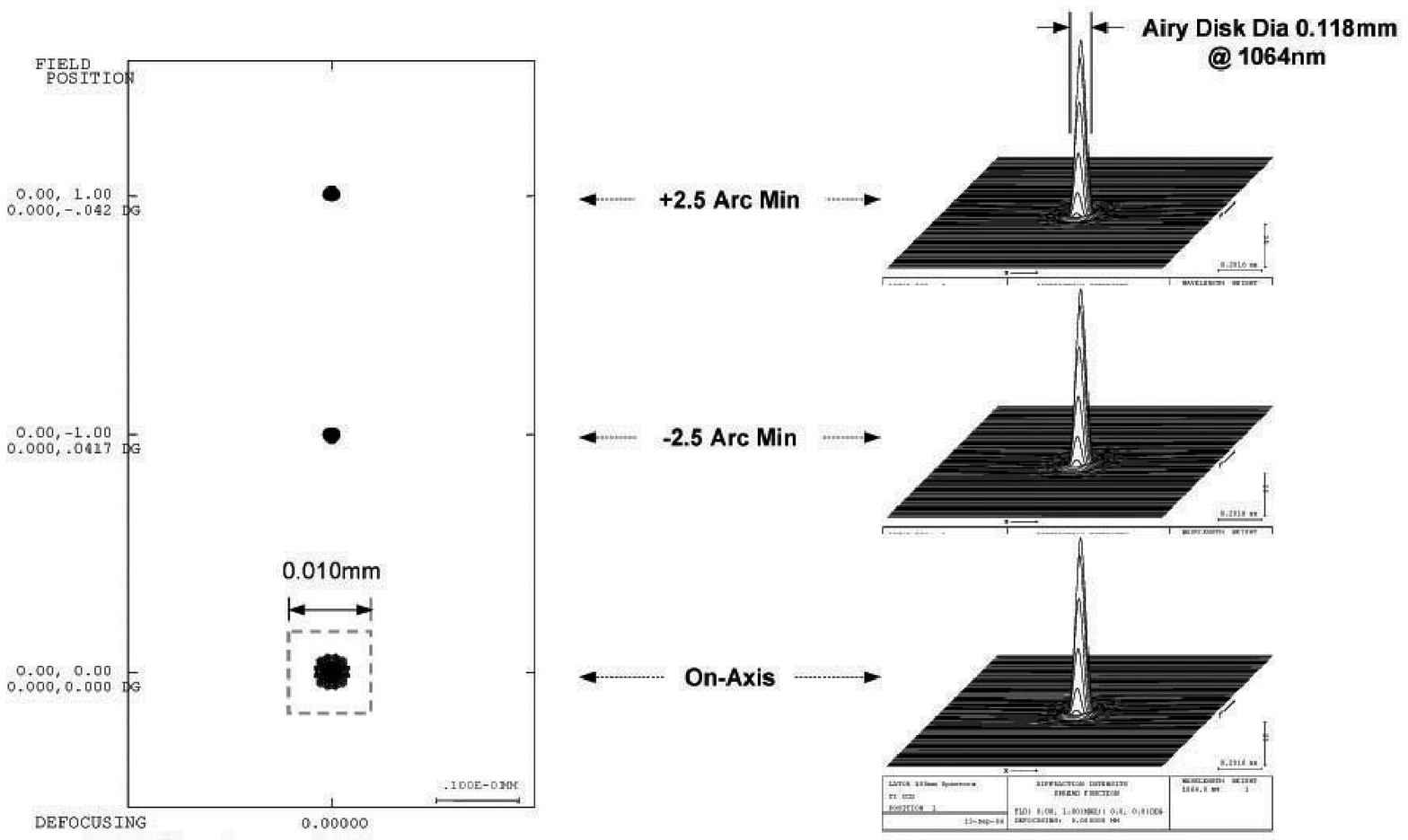,width=158mm}%,height=90mm}
\end{center}
\vskip -10pt 
  \caption{CCD channel geometric (left) and diffraction (right) PSF.  
 \label{fig:lator_ccd}}
\end{figure*}

Figure \ref{fig:lator_focal} shows the design of the focal plane capping. The straight edge of the `D'-shaped CCD field stop is tangent to the limb of the Sun and it is also tangent to the edge of APD field stop (pinhole). There is a 2.68 arcsecond offset between the straight edge and the concentric point for the circular edge of the CCD field stop (`D'-shaped aperture). In addition, the APD field of view and the CCD field of view circular edges are concentric with each other. Depending on the spacecraft orientation and pointing ability, the `D'-shaped CCD field stop aperture may need to be able to be rotated to bring the straight edge into a tangent position relative to the limb of the Sun. 
The results of the analysis of APD and CCD channels point spread functions (PSF) are shown in Figures \ref{fig:lator_apd} and \ref{fig:lator_ccd}.

%************
%**************

%************
%**************

\subsubsection{The Factors Affecting Signal to Noise Analysis}

In conducting the signal-to-noise analysis we pay significant attention to several important factors. In particular, we estimate what fraction of the transmitted signal power captured by the 100 mm receiver aperture and analyze the effect of the Gaussian beam divergence (estimated at $\sim  7 ~\mu$rad) of the 200 mm transmit aperture on the ISS. Given the fact that the distance between the transmitter and receiver is on the order of 2 AU, the amount captured is about $2.3\times 10^{-10}$ of the transmitted power.  

We also consider the amount of solar disk radiation scattered into the two receiver focal planes. In particular, the surface contamination, coating defects, optical roughness and substrate defects could scatter as much as $1\times10^{-4}$ or more (possibly $1\times 10^{-3}$)  of the solar energy incident on the receive aperture into the field of view.   These issues are being considered in our current analysis. We also study the amount of the solar corona spectrum within the receive field of view that is not blocked by the narrow band pass filter.  The factors we consider is the filter's FWHM band-pass is 2 nm, the filter will have 4.0 optical density (OD) blocking outside the 2 nm filter band pass from the X-ray region of 1200 nm; the filter efficiency within the band pass will be about 35\%, and the detector is probably sensitive from 300 nm to 1200 nm.  

Additionally, we consider the amount of out-of-field solar radiation scattered into the focal plane by the optical housing. This issue needs to be investigated in a stray light analysis which can be used to optimize the baffle design to minimize the stray light at the focal plane.  Finally, we study the effectiveness of the baffle design in suppressing stray light at the focal plane. Thus, in addition to the stray light analysis, the effectiveness of the final baffle design should be verified by building an engineering model that can be tested for stray light.  

Our recent conceptual design and a CODEV  ray-trace analysis met all the configuration and performance requirements (shown in Table \ref{table:requirements}).  The ray-trace performance of the resulted instrument is diffraction limited in both the APD and CCD channels over the specified field of view at 1064 nm. The design incorporated the required field stop and Layot stop. A preliminary baffle design has been developed for controlling the stray light.  
In the near future, we plan to perform a stray light analysis which should be performed to optimize the baffle design and calculate the amount of stray light that could be present at each of the two focal planes.  This stray light analysis will take into account the optical smoothness of the band-pass filter and primary mirror surfaces. Narrow angle scattering may be a problem at the two focal planes in the filter and primary mirror are not optically very smooth and, thus, it requires a more detailed study. Finally, a rigorous signal-to-noise analysis will be performed to validate the power required to achieve a high signal-to-noise ratio in detecting received beam signal in the presence of the expected focal beam stray light predicted by the stray light analysis and the engineering model stray light tests.

The importance of this design is in the fact that it can be applied for many applications, thus, opening new ways for optical communication, accuracy navigational and fundamental physics experiments. This LATOR-related design experience motivates us to think about an architecture that may have a much border uses for the purposes of precision navigation and high data rate transmission and capable to operate at large interplanetary distance. In the next section we will summarize our current ideas.   

\section{A Multi-Purpose Space Architecture}
\label{sec:tech}

The three different applications presented in the sections above highlight the scientific potential of laser ranging applications from ranging to the Moon, to Mars, and over interplanetary distances. Thus, LLR could reach an order of magnitude improvements in the tests of relativistic gravity once the APOLLO facility becomes operational. A laser transponder on Mars would test the SEP violation parameter $\eta$ with accuracy of $\sim 5\times 10^{-6}$. Such a experiment may provide a measurement of the Eddington parameter  $\gamma$ to $8\times 10^{-7}$, and parameter $\beta$ to $\sim 1.25 \times 10^{-5}$. It would also improve the test of temporal variation of the gravitational constant to accuracy of $\dot{G}/G\sim 10^{-14}$ yr$^{-1}$. The LATOR mission that uses inter-spacecraft laser ranging at distances of $\sim2$ AU would improve the measurement of the Eddington parameter $\gamma$ to $\sim 1\times 10^{-8}$ in a single measurement. How we can enable these missions? What technologies are needed to provide such an unprecedented accuracy leap? A move from the currently standard microwave communication technologies to the communication and navigation at the optical frequencies is the natural step.  
What would it take to build an optical system that would be able to
perform in all these different environments? In this section we are trying to answer these questions and to propose a multi-purpose architecture that would enable even more accurate tests of fundamental physics, while providing for the goals of solar system exploration.

\subsection{An Optical Architecture for Deep Space Exploration}

Prohibitively large cost of dedicated missions motivate us to consider development of a system with the possibility of dual application -- for both exploration and fundamental physics research. A recent  progress in the design of the optical communication system for the interplanetary opportunities promises very exciting developments in this area. The first interplanetary lasercomm system will be deployed with the NASA Mars Telesat Orbiter mission that is readying for launch in 2009. With the flight terminal being built by MIT Lincoln Laboratory and the ground terminals being built and operated by JPL, the NASA-run program hopes to gain knowledge
and experience useful for designing, procuring, and operating cost-effective future deep-space optical communication systems. Such systems are predicted to be able to provide the 1-3 orders of magnitude higher data rates required by future science and exploration
missions.  This mission, as one of its goals, has an optical link demonstration that would use the 1.6 m telescope on Table Mountain as the transmitter system and the Palomar 5 m telescope as the receiver. A clock, a time tagging unit and a photodetection system are implemented on board the spatial vehicle that will orbit around Mars. The corresponding experiment called the Mars Laser Communication Demonstration (MLCD) experiment \cite{Boroson_etal_2004}. 

The principle of the MLCD experiment lies on the measure of the propagation duration of laser pulses emitted from an Earth laser station in the Martian orbiter direction. These laser pulses are timed on Earth and on board the orbiter, respectively in the time scale of the terrestrial clock and in the time scale of the orbiter clock. The distance between the Earth and the orbiter is computed from the difference between the start time and the arrival time on board. Since the link budget depends on the  distance as $d^{-2}$ as compared $d^{-4}$ for the two-ways laser ranging, measurements in the range of the solar system can be envisioned. Such a laser ranging telemetry has never been realized and has many technological and scientific applications. 

The hardware being built for the MLCD may become a critical component of the new space multi-purpose architecture. The components of a new architecture for deep space exploration will rely on the laser-ranging techniques, optical transmitters and receivers, optical transponders, precision clocks---and possibly SQUIDs. This new optical infrastructure would allow one to address the goals of several physics areas and would also lead to a multi-disciplinary approach in the areas of gravitation and fundamental physics, materials science, lunar and planetary interiors and geology, and Mars orbital dynamics and interior structure. In addition, such an architecture may be very useful in a number of space exploration themes, namely 3D-sensing, altimetry, navigation, attitude control, communication, precision alignment as needed for construction, power generation and distribution, in-space structural control, in-space laser propulsion, resource location and many others.  As such this new system should be constructed for the benefits of science and exploration.

\subsection{Desired Capabilities}

The concept for the all-in-space optical architecture should wrap around two critical capabilities---optical communication and navigation. As such, it has several critical subsystems, consisting of a laser, a clock, receiver/optics, an APD array, and capability for significant articulation. It should be of small size and light-weight.  Finally a system like that should be a general purpose communication/navigation unit and could be used for a possible deployment on the surfaces of solar system bodies, Moon, Mars, Phobos, Deimos, Mercury, Jupiter's icy moons; it may be integrated with spacecraft, used on the planetary bases, and on the ISS.

The receiver location is currently one of the primary limitations of an all-optical communication system.  For an architecture like that to become a mainstream system for deep space application, it must be located above the Earth's atmosphere. Among the possible receivers locations one may consider putting receivers on the Moon, or deploy them at Earth-Moon Lagrange (EL2) point.  The system can be placed on a constellation of LEO stations with an RF link to Earth, which could help to beat atmospheric effects. The space deployment solves the ``last-mile'' ultra-broad-band deep-space communication problem, which needs an RF data link with the Earth-based system.

As for the expected capabilities for such an infrastructure, we expect
that existing technologies can provide for a construction of a
system with laser  wavelength of 1064 nm, power of 1-2~W, receiver aperture $\sim$ 20 cm. For communication purposes, such a system can  achieve a high data rate of $\sim$ few Gbps over 2 AU distances. The system would have built-in networking capabilities with
the use of multiple frequencies [i.e 532 nm / 1064 nm as an option]. This
system would initially operate in a photon-counting mode, which is quite
understood given the significant expertise gained with lunar
and satellite laser ranging. However, if a phase-referencing capabilities are required (for a coherent communication, for instance), the system would need to rely on a precision clock. The use of an acceleration compensation, a drag free-environment (in free-space) or laser metrology could further improve its performance.

To date, the most precise ranging to another planet was achieved when the radio telescopes of the NASA/JPL Deep Space Network (DSN) were in communication with the Viking Lander on Mars. This two-way microwave link typically resulted in several meter precision measurements, with 3 meters precision being the best internal consistency ever reported over a short tracking interval. The absolute accuracy of these interplanetary measurements, however, is limited by the presence of the interplanetary solar plasma, which, like the Earth's similarly charged ionosphere, can have a significant, and largely unknown, impact on the propagation delay at microwave frequencies. Optical frequencies, on the other hand, are much too high to be affected by the charged solar plasma. Furthermore, the shorter optical wavelengths characteristic of lasers allow the transmitted energy to be propagated in highly collimated (few arcsecond divergence) beams and therefore require much smaller collecting and transmitting apertures at both ends of the link. Of course, mutual high accuracy pointing between the space and ground terminals is required to take advantage of the narrow laser beam divergence. Furthermore, because of the long pulse transit times between terminals, each of the remote
terminals must be able to independently detect, identify, and lock onto the signals from the opposite terminal using only a priori information stored at that terminal.  

As far as deep-space navigation is concerned, one may expect
cm-class precision for spacecraft at distances up to 5 AU.  The frequency modulation will be critical for the systems that utilize CW lasers for the communications purposes. The bandwidth for the modulation of the carrier signal will be of most importance. 
To achieve such a level of navigation, a significant effort is required to fully deploy this option. The Mars Telesat Orbiter is the first opportunity; additional investments are needed to develop the next steps. Since such single-ended systems
are incapable of ranging beyond the Moon to the planets, we consider the feasibility of a two-way asynchronous (i.e. independently firing) interplanetary laser transponder pair, capable of decimeter ranging and subnanosecond time transfer from Earth to a spacecraft anywhere within the inner solar system \cite{Degnan_2002}. Therefore, the development of the asynchronous optical transponder systems is a critical effort for the future precise navigation in the deep space.

Design of such a system for interplanetary use presents additional  challenges. Thus, working in the solar proximity may present some complications. What would be possible solutions to counter-play the effect of solar proximity?  A natural solution would be to have an accurate navigation and attitude information on the spacecraft in addition to the solar avoidance software. One would have to choose operating wavelength with good atmospheric transmission far from the solar blackbody peak.  A solution to this would be the use of diode-pumped,  sub-nanosecond pulse, Er:YAG microchip lasers and fiber amplifiers that emit at 1550 nm where there is excellent atmospheric transmission.   One would also need a fast photon-counting technology that exist but has relatively low quantum efficiency.  

Similar to the optical design of the LATOR mission (presented in the next Section \ref{sec:lator}), one can design a system capable to reflect most or all solar irradiance from telescope surfaces and entrance window. Alternative solutions include the use of `cold mirror' window that reflects the visible and transmits the infrared or `white' the surfaces on the telescope exterior to reflect most of the solar radiation. Again, similar to the LATOR design, one can filter out residual visible/NIR light by using primary and secondary `hot mirrors' that reflect infrared and transmit residual visible light to heat-sinked absorbers decoupled from the optical bench.  The common design considerations for most of the optical systems for the interplanetary applications is the aggressive stray light control, the use of the light-tight receiver boxes and/or light tubes, the application of the narrow spatial field of view and narrow bandpass filters.

The outlined lasercomm concept relies on the exisitng technology developed to for precise spatial acquisition, tracking and fine beam-pointing.  The  instrument will be capable of signal acquisition on a noisy background and will benefit from vibration isolation technology. One may consider building a lasercomm station on the ISS or the Moon, which would be the primary infrastructure for deep space navigation and communication over interplanetary distances. Such an infrastructure would not only demonstrate benefits of this technology for space exploration, but would also allow precision tests of fundamental and gravitational physics. 

\section*{Conclusion}

Lasers are contributing to a wide field of astrophysics. Thus, a Nd:YAG laser at 10.6 $\mu$m wavelength will be used in the first experiments to attempt detecting gravitational waves from sources like supernovas and orbiting neutron stars. These experiments use interferometers that should be capable of measuring a change in length
between the two interferometer arms to a precision of one part in 10$^{22}$. A space warp of this magnitude is predicted for gravitational radiation from astrophysical sources. The terrestrial experiments are called LIGO (Light Interferometer Gravitational Wave Observatory) in the U.S. and GEO in Europe. A space-based experiment
called LISA (Light Interferometer Space Antenna) is also in progress. The LIGO interferometer arms are each 4 km long. A frequency-stable, low noise, high-spatial-beam-quality laser at a power level of 10 W is required for the light source. Cavity mirrors form resonators in each interferometer arm that increase the power in the cavities to nearly 1 kW. Four Nd:YAG rods, each side pumped by two 20 W diode bars, amplify the single frequency output of a non-planar ring oscillator from 700 mW to at least 10 W. Achieving the required sensitivity for detecting gravitational waves means resolving each interferometer fringe to one part in 10$^{11}$, a formidable, but an achievable goal.

The laser ranging technology is important for space exploration! The progress alone this road would lead us to development of new technologies, practical applications. What is important is that we need to further test relativistic gravity as gravitation is the most elusive force of nature.  The expected cm accuracy ranging over interplanetary distances could give spectacular scientific results.  The recent activities in regards to the ''Moon-Mars Initiative'' have already motivated  development of consolidated prediction formats and procedures for a variety of laser ranging targets, including: i) the lunar prediction formats and software for possible new lunar ranging stations; ii) the anticipated laser transponders on the lunar surface or on orbit about the Moon; iii)  the ongoing technology development for transponders to be deployed on the Martian surface and other solar system bodies, etc. These are the current activities undertaken aimed to improve the current predictions for the laser ranging in the near future.

There are many activities that focused on the design on an optical transponder which will work at the distance of Mars and test it at the Moon. This paper summarizes required capabilities for such a system.   Optical communication is planned for future NASA interplanetary missions.  An optical communication demonstration is planned for NASA 2009 Mars Telesat Orbiter. LLR generates strong, and often unique, science results that should be continued in the future. The APOLLO facility will provide new opportunities for fundamental physics enabled by new optical technologies. The proposed LATOR mission would take us even further. 

Developing laser technologies and their contributions to science are too numerous to cover adequately in this article. Laser communications between satellite networks, laser propelled spacecraft and laser fusion are additional examples of developing laser technologies. There are many new fundamental physics experiments that
are being enabled by laser technology including correction for atmospheric distortions in astronomy using laser reflections from the sodium layer in the upper atmosphere and studies of quantum electrodynamics using ultra-intense laser beams. Just as it was hard to envision the potential of laser technologies in the 1960s and
1970s, it seems clear that we cannot now envision the many new developments in lasers and their applications in the next century will see.  The present time requires an imaginative and ambitious approach, the discussed multi-purpose lasercomm system is the logical step in this direction.  

%**************************************
\subsubsection*{Acknowledgments~~} 
The work described here was carried out at the Jet Propulsion Laboratory, California Institute of Technology, under a contract with the National Aeronautics and Space Administration.

%**************************************

%***************************

\begin{thebibliography}{99}

\bibitem%[Adelberger(2001)]
{Adelberger_2001} Adelberger, E. G., ``New Tests of Einstein's Equivalence Principle and Newton's inverse-square law,'' 
{\it Classical and Quantum Gravity \bf 18}, 2397-2405 (2001).

\bibitem%[Allen(2000)]
{Allen_2000} Allen, C. W. ``Astrophysical Quantities'' (AIP Press, 4th edition, 2000).

\bibitem%[Anderson et al.(1996)]
{Anderson_etal_1996} Anderson, J.~D., Gross, M.,  Nordtvedt, K.~L., and Turyshev, S.~G., ``The Solar Test of the Equivalence Principle,'' {\it Astrophysical Journal \bf 459}, 365-370 (1996).

\bibitem%[Anderson and Williams(2001)]
{Anderson_Williams_2001} Anderson, J. D., and Williams, J. G., ``Long-Range Tests of the Equivalence Principle,'' {\it Classical and Quantum Gravity \bf 18}, 2447-2456 (2001).

\bibitem%[Anderson et al.(2002)]
{Anderson_etal_2002}  
Anderson, J.~D., Lau, E. L., Turyshev, S. G.,  Williams, J. G.,  Nieto, M. M., ``Recent Results for Solar-System Tests of General Relativity.'' Presented at 200-th AAS Meeting, Albuquerque, NM (2-6 June 2002).  Paper \#12.06, {\it BAAS \bf 34}, 833 (2002).

\bibitem%[Anderson, Lau, and Giampieri.(2004)]
{Anderson_Lau_Giampieri_2004} Anderson, J. D., Lau, E. L., Giampieri, G., ''Improved Test of General Relativity with Radio Doppler Data from the Cassini Spacecraft.'' In Proc. ``The XXII Texas Symposium on Relativistic Astrophysics,'' Stanford University, December 13-17, 2004,  to be published.

\bibitem%[Baessler et al.(1999)]
{Baessler_etal_1999} Baessler, S., Heckel, B., Adelberger, E., Gundlach, J., Schmidt, U., and Swanson, E., "Improved Test of the Equivalence Principle for Gravitational Self-Energy," {\it Phys. Rev. Letters \bf 83}, 3585-3588 (1999).  

\bibitem%[Bertotti, Iess, and Tortora(2003)]
{Bertotti_Iess_Tortora_2003} Bertotti, B., Iess, L., Tortora, P., ``A test of general relativity using radio links with the Cassini spacecraft," {\it Nature \bf 425}, 374-376 (2003).

\bibitem%[Bender et al.(1973)]
{Bender_etal_1973} Bender, P. L., Currie, D. C., Dicke, R. H., Eckhardt, D. H., Faller, J. E., Kaula, W. M., Mulholand, J. D., Plotkin, H. H., Poultney, S. K., Silverberg, E. C., Wilkinson, D. T., Williams, J. G. \& Alley, C. O., ``The Lunar Laser Ranging Experiment,'' {\it Science \bf 182}, 229-237 (1973).

\bibitem%[Boroson et al.(2004)]
{Boroson_etal_2004} Boroson, D.~M., Biswas, A., Edwards, B.~ L., ``MLCD: Overview of NASA's Mars Laser Communications Demonstration System.'' In proc. ``Free-Space Laser Communication Technologies XVI,'' edited by G. Stephen Mecherle, Cynthia Y. Young, John S. Stryjewski, (SPIE, Bellingham, WA, 2004). {\it Proc. of SPIE vol. \bf 5338}, 16-28 (2004).

\bibitem%[Chandler et al.(1994)]
{Chandler_etal_1994} Chandler, J. F.,  Reasenberg, R. D.,  Shapiro I. I., ``New Results on the Principle of Equivalence.''
{\it BAAS \bf 26},  1019 (1994).

\bibitem%[Damour(1996)]
{Damour_1996} Damour, T., ``Testing the Equivalence Principle: why and how?'' {\it Classical and Quantum Gravity \bf 13}, A33 (1996).

\bibitem%[Damour(2001)]
{Damour_2001} Damour, T., ``Questioning the Equivalence Principle?'' [arXiv:gr-qc/0109063].

\bibitem%[Damour and Esposito-Farese(1996)]
{Damour_EFarese96} 
Damour, T., Esposito-Farese, G., ``Testing gravity to second post-Newtonian order: A field-theory approach,'' {\it Phys. Rev. D \bf 53}, 5541-5578 (1996). 

\bibitem%[Damour and Nordtvedt(1993a)]
{Damour_Nordtvedt_1993a} 
	Damour, T., Nordtvedt, K. Jr., ``General Relativity as a Cosmological Attractor of Tensor Scalar Theories'', {\it Phys. Rev. Letters \bf 70} 2217-2219 (1993).

\bibitem%[Damour and Nordtvedt(1993b)]
{Damour_Nordtvedt_1993b} Damour, T., Nordtvedt, K., ``Tensor-scalar cosmological models and their relaxation toward general relativity,'' {\it Phys. Rev. D \bf 48}, 3436-3450 (1993).

\bibitem%[Damour and Polyakov(1994a)]
{Damour_Polyakov_1994a} Damour, T., and Polyakov, A. M., ``String Theory and Gravity,'' {\it Gen. Relativ. Grav. \bf 26}, 1171-1176 (1994a).

\bibitem%[Damour and Polyakov(1994b)]
{Damour_Polyakov_1994b} Damour, T., and Polyakov, A. M., ``The string dilaton and a least coupling principle,'' {\it Nucl. Phys. B \bf423}, 532-558 (1994b).

\bibitem%[Damour, Piazza, and Veneziano (2002a)]
{Damour_Piazza_Veneziano_2002a} 
Damour, T., Piazza, F., Veneziano, G., ``Runaway dilaton and equivalence principle violations,'' {\it Phys. Rev. Letters \bf 89} 081601 (2002), [arXiv:gr-qc/0204094].

\bibitem%[Damour, Piazza, and Veneziano (2002b)]
{Damour_Piazza_Veneziano_2002b} 
Damour, T., Piazza, F., Veneziano, G., ``Violations of the equivalence principle in a dilaton-runaway scenario,''
{\it Phys. Rev. D \bf 66} 046007 (2002), [arXiv:hep-th/0205111].

\bibitem%[Damour and Vokrouhlicky (1996a)]
{Damour_Vokrouhlicky_1996a} Damour, T., and Vokrouhlicky, D., ``Equivalence Principle and the Moon,'' {\it Phys. Rev. D \bf 53}, 4177-4201 (1996a).  

\bibitem%[Damour and Vokrouhlicky (1996b)]
{Damour_Vokrouhlicky_1996b} Damour, T., and Vokrouhlicky, D., ``Testing for gravitationally preferred directions using the lunar orbit,'' {\it Phys. Rev. D \bf 53}, 6740-6740 (1996b).  

\bibitem%[Dickey, Newhall, and Williams(1989)]
{Dickey_etal_1989} Dickey, J. O., Newhall, X X, and Williams, J. G., ``Investigating Relativity Using Lunar Laser Ranging: Geodetic Precession and the Nordtvedt Effect,'' {\it Adv. Space Res. \bf 9}, 75-78 (1989).

\bibitem%[Dickey et al.(1994)]
{Dickey_etal_1994} Dickey, J. O., Bender, P. L., Faller, J. E., Newhall, X X, Ricklefs, R. K., Shelus, P. J., Veillet, C., Whipple, A. L., Wiant, J. R., Williams, J. G., and Yoder, C. F.: ``Lunar Laser Ranging: A Continuing Legacy of the Apollo Program,'' {\it Science \bf 265}, 482-490 (1994).

\bibitem%[Degnan(2002)]
{Degnan_2002} Degnan, J.~J., ``Asynchronous laser transponders for precise interplanetary ranging and time transfer,'' {\it Journal of Geodynamics \bf 34}, 551–594 (2002).

\bibitem%[Marciano(1984)]
{Marciano1984} Marciano, W.~J., ``Time Variation of the Fundamental "Constants" and Kaluza-Klein Theories,'' {\it Phys. Rev. Letters \bf 52}, 489-491 (1984).

\bibitem%[Moyer(1977)]
{Moyer_1977} Moyer, T.~D., JPL Internal Memorandum No. 314.7-122, 1977, unpublished.

\bibitem%[Moyer(2003)]
{Moyer_2003} Moyer, T.~D., ``Formulation for Observed and Computed Values of Deep Space Network Data Types for Navigation'' (Willey, 2003) 

\bibitem%[Murphy et al.(2002)]
{Murphy_TM_etal_2002} Murphy, T. M., Jr., Strasburg, J. D., Stubbs, C. W., Adelberger, E. G., Angle, J., Nordtvedt, K., Williams, J. G., Dickey, J. O., \& Gillespie, B., ``The Apache Point Observatory Lunar Laser-Ranging Operation (APOLLO),'' Proceedings of 12th International Workshop on Laser, Ranging, Matera, Italy, November 2000, in press, (2002),  {\tt
http://www.astro.washington.edu/tmurphy/apollo/matera.pdf}

\bibitem%[Nordtvedt(1968a)]
{Nordtvedt_1968a} Nordtvedt, K., Jr., ``Equivalence Principle for Massive Bodies.  I. Phenomenology,'' {\it Phys. Rev. \bf 169}, 1014–1016 (1968a).

\bibitem%[Nordtvedt(1968b)]
{Nordtvedt_1968b} Nordtvedt, K., Jr., ``Equivalence Principle for Massive Bodies. II. Theory.'' {\it Phys. Rev. \bf 169}, 1017–1025 (1968b).

\bibitem%[Nordtvedt(1968c)]
{Nordtvedt_1968c} Nordtvedt, K., Jr., ``Testing Relativity with Laser Ranging to the Moon,'' {\it Phys. Rev. \bf 170}, 1186–1187 (1968c).

\bibitem%[Nordtvedt(1970)]
{Nordtvedt_1970} Nordtvedt, K., Jr., ``Solar system 
E\"otvos experiments,'' {\it Icarus \bf 12}, 91-100, (1970). 

\bibitem%[Nordtvedt(1987)]
{Ken_2PPN_87}	
Nordtvedt, K., ``Probing Gravity to the 2nd Post-Newtonian Order and to one part in 10$^7$ Using the Sun,'' {\it Astrophysical Journal \bf 320}, 871-874 (1987). 

\bibitem%[Nordtvedt(1991)]
{Nordtvedt_1991} Nordtvedt, K., Jr., ``Lunar Laser Ranging Re-examined: The Non-Null Relativistic Contribution,'' {\it Phys. Rev. D \bf 43}, 3131-3135 (1991).

\bibitem%[Nordtvedt(1998)]
{Nordtvedt_1998} Nordtvedt, K., ``Optimizing the observation schedule for tests of gravity in lunar laser ranging and similar experiments'', {\it Classical and Quantum Gravity \bf 15}, 3363-3381 (1998).  

\bibitem%[Nordtvedt(1999)]
{Nordtvedt_1999} Nordtvedt, K., ``30 years of lunar laser ranging and the gravitational interaction,'' {\it Classical and Quantum Gravity \bf 16}, A101-A112 (1999).  

\bibitem%[Nordtvedt, M\"uller, and Soffel(1995)]
{Nordtvedt_Muller_Soffel_1995} Nordtvedt, K.~L., M\"uller, J., Soffel, M., ``Cosmic Acceleration of the Earth and Moon by Dark-Matter,'' {\it Astron. and Astrophys. \bf 293}, L73-L74 (1995).  

\bibitem%[Nordtvedt(2003)]
{Nordtvedt_2003}	
Nordtvedt, K., Jr., ``Lunar Laser Ranging - A Comprehensive Probe of Post-Newtonian Gravity,'' [arXiv:gr-qc/0301024].

\bibitem%[Holdridge(1967)]
{Holdridge_1967} Holdridge, D.~B., in {\it Supporting Research and Advanced Development}, Space Programs Summary 37–48, Jet Propulsion Laboratory report, unpublished, Vol. III, pp. 2–4 (1967).

\bibitem%[Hellings et al.(1983)]
{Hellings_etal_1983} Hellings, R. W., P. J. Adams, J. D. Anderson, M. S. Keesey, E. L. Lau, E. M. Standish, V. M. Canuto, and I. Goldman, ``Experimental Test of the Variability of G Using Viking Lander Ranging Data,'' {\it Phys. Rev. Lett. \bf 51}, 1609-1612 (1983).

\bibitem%[Hellings(1983)]
{Hellings_1983} Hellings, R. W., ``Testing General Relativity with Solar System Dynamics,'' in Proceedings of the Tenth International Conference on General Relativity and Gravitation, B. Bertotti, ed., North Holland (1983).

\bibitem%[Hellings(1984)]
{Hellings_1984} Hellings, R. W., ``Testing Relativity with Solar System Dynamics,'' in {\it General Relativity and Gravitation}, B. Bertotti, F. de Felice, and A. Pascolini, eds., pp. 365-385, Reidel, 1984.

\bibitem%[Reasenberg et al.(1979)]
{Reasenberg_etal_1979} Reasenberg, R. D., Shapiro, I. I., MacNeil, P. E., Goldstein, R. B., Breidenthal, J. C., Brenkle, J. P., Cain, D. L., Kaufman, T. M., Komarek, T. A., Zygielbaum, A. I., ``Viking relativity experiment - Verification of signal retardation by solar gravity,''  {\it Astrophysical Journal \bf 234}, L219-L221 (1979).

\bibitem%[Robertson, Carter, and Dillinger(1991)]
{Robertson_etal_1991} Robertson, D. S., Carter, W. E., and Dillinger, W. H., ``A New Measurement of Solar Gravitational Deflection of Radio Signals Using VLBI,'' {\it Nature \bf 349}, 768-770 (1991).

\bibitem%[Shapiro(1964)]
{Shapiro_1964} Shapiro, I.~I., ``Fourth Test of General Relativity,''{\it Phys. Rev. Lett. \bf 13}, 789-791 (1964).

\bibitem%[Shapiro, Counselman, and King(1976)]
{Shapiro_Counselman_King_1976} Shapiro, I. I., Counselman, C. C., III, and King, R. W.: ``Verification of the Principle of Equivalence for Massive Bodies,'' {\it Phys. Rev. Lett. \bf 36}, 555-558 (1976).

\bibitem%[Shapiro et al.(1977)]
{Shapiro_etal_1977} Shapiro, I. I., Reasenberg, R. D., MacNeil, P. E., Goldstein, R. B., Brenkle, J. P., Cain, D. L., Komarek, T., Zygielbaum, A. I., Cuddihy, W. F., and Michael, W. H., Jr., ``The Viking Relativity Experiment,'' {\it J. Geophys. Res. \bf 82}, 4329-4334 (1977).

\bibitem%[Shapiro et al.(1988)]
{Shapiro_etal_1988} Shapiro, I. I., Reasenberg, R. D., Chandler, J. F., and Babcock, R. W., ``Measurement of the de Sitter Precession of the Moon: A Relativistic Three-Body Effect,'' {\it Phys. Rev. Lett. \bf 61}, 2643-2646 (1988).

\bibitem%[Shapiro(1990)]
{Shapiro_1990} Shapiro, I. I., ``Solar System Tests of General Relativity: Recent Results and Present Plans,'' in Proceedings of the 12th International Conference on General Relativity and Gravitation, University of Colorado at Boulder, July 2-8, 1989, editors N. Ashby, D.~F. Bartlett, and W. Wyss, Cambridge (1990).

\bibitem%[Shapiro et al.(2003)]
{Shapiro_SS_etal_2004}
Shapiro, S.~S., Davis, J.~L., Lebach, D.~E., and  Gregory, J.~S.,
''Measurement of the Solar Gravitational Deflection of RadioWaves using Geodetic Very-Long-Baseline Interferometry Data, 1979–1999,''
{\it Phys. Rev. Lett. \bf 92}, 121101 (2004).

\bibitem%[Spergel et al.(2003)]
{Spergel:2003cb}
Spergel, D.~N., Verde, L., Peiris, H.~V., Komatsu, E., Nolta, M.~R., Bennett, C.~L., Halpern, M., Hinshaw, G., Jarosik, N., Kogut, A., Limon, M., Meyer, S.~S., Page, L., Tucker, G.~S., Weiland, J.~L., Wollack, E., Wright, E.~L.,
``First Year Wilkinson Microwave Anisotropy Probe (WMAP) Observations: Determination of Cosmological Parameters,''
{\it Astrophys.\ J.\ Suppl.\  \bf 148}, 175 (2003), [arXiv:astro-ph/0302209]. 

\bibitem%[Turyshev et al.(2004)]
{Turyshev_etal_2004} Turyshev, S.~G., Williams, J.~G., Nordtvedt, K., Shao, M., Murphy, T.~W. Jr., ``35 Years of Testing Relativistic Gravity: Where do we go from here?'' In Proc. ``302.WE-Heraeus-Seminar: Astrophysics, Clocks and Fundamental Constants, 16-18 June 2003. The Physikzentrum, Bad Honnef,  Germany,'' Springer Verlag, {\it Lect. Notes Phys.  \bf 648}, 301-320 (2004), [arXiv:gr-qc/0311039].

\bibitem%[Turyshev, Shao, and Nordtvedt(2004)]
{lator_cqg2004}  
Turyshev, S.~G., Shao, M., \& Nordtvedt, K.~L.,
{``The Laser Astrometric Test of Relativity Mission''},   
{\it Classical and Quantum Gravity \bf 21}, 2773-2799, 2004, [arXiv:gr-qc/0311020].

\bibitem%[Ulrich(1982)]
{Ulrich_1982} Ulrich, R. K., ``The Influence of Partial Ionization and Scattering States on the Solar Interior Structure,'' {\it Astrophysical Journal \bf 258}, 404-413 (1982).

\bibitem%[Will(1971)]
{Will_1971} Will, C.M., ``Theoretical frameworks for testing relativistic gravity. III. Conservation laws, Lorentz invariance, and values of the PPN parameters,''{\it Astrophysical Journal \bf 169}, 125-140 (1971).

\bibitem%[Will and Nordtvedt(1972)]
{Will_Nordtvedt_1972} Will, C. M. and Nordtvedt, K., ``Conservation Laws and Preferred Frames in Relativistic Gravity I: Preferred-Frame Theories and an Extended PPN Formalism,'' {\it Astrophysical Journal \bf 177}, 757-774 (1972).

\bibitem%[Will(1990)]
{Will_1990} Will, C.~M., ``General Relativity at 75: How Right was Einstein?'' {\it Science \bf 250}, 770-771 (1990).

\bibitem%[Will(1993)]
{Will_1993} Will, C. M.,  ``Theory and Experiment in Gravitational Physics,'' (Cambridge University Press), 1993. 

\bibitem%[Will(2001)]
{Will_2001} Will, C.~M., ``The Confrontation between General Relativity and Experiment,'' {\it Living Rev. Rel. \bf 4}, 4 (2001), [arXiv:gr-qc/0103036].  

\bibitem%[Williams et al.(1976)]
{Williams_etal_1976} Williams, J. G., Dicke, R. H., Bender, P. L., Alley, C. O., Carter, W. E., Currie, D. G., Eckhardt, D. H., Faller, J. E., Kaula, W. M., Mulholland, J. D., Plotkin, H. H., Poultney, S. K., Shelus, P. J., Silverberg, E. C., Sinclair, W., S., Slade, M. A., and Wilkinson, D. T., ``New Test of the Equivalence Principle from Lunar Laser Ranging,'' {\it Phys. Rev. Lett. \bf 36}, 551 (1976).

\bibitem%[Williams, Newhall, and Dickey(1996a)]
{Williams_Newhall_Dickey_1996a} Williams, J.~G., Newhall, X~X, and Dickey, J.~O., ``Relativity Parameters Determined from Lunar Laser Ranging,'' {\it Phys. Rev. D \bf 53}, 6730-6739 (1996a).

\bibitem%[Williams, Newhall, and Dickey(1996b)]
{Williams_Newhall_Dickey_1996b} Williams, J.~G., Newhall, X~X., and J.~O. Dickey, ``Relativity parameters determined from lunar laser ranging,'' The Seventh Marcel Grossmann meeting {\it On recent developments in theoretical and experimental general relativity, gravitation, and relativistic field theories}. Proceedings of the meeting held at Stanford University, 24-30 July 1994, editors R.~T. Jantzen and G. Mac Keiser, World Scientific, Singapore, 1529-1530 (1996b).

\bibitem%[Williams et al.(2002)]
{Williams_etal_2002} Williams, J.~G., Boggs, D.~H., Dickey, J.~O., and Folkner, W.~M., ``Lunar Laser Tests of Gravitational Physics,'' in proceedings of The Ninth Marcel Grossmann Meeting, World Scientific Publ., editors V. G. Gurzadyan, R. T. Jantzen, and R. Ruffini, 1797 (2002).

\bibitem%[Williams and Dickey(2003)]
{Williams_Dickey_2003} Williams, J.~G. and Dickey, J.~O., ``Lunar Geophysics, Geodesy, and Dynamics,'' proceedings of 13th International Workshop on Laser Ranging, October 7-11, 2002, Washington, D. C. (2003) {\tt http://cddisa.gsfc.nasa.gov/lw13/lw$\underline{}$proceedings.html}

\bibitem%[Williams, Turyshev, and Murphy(2004)]
{Williams_Turyshev_Murphy_2004} Williams, J.~G., Turyshev, S.~G., Murphy, T.~W., Jr., ``Improving LLR Tests of Gravitational Theory.'' {\it International Journal of Modern Physics D \bf 13}, 567-582 (2004), [arXiv:gr-qc/0311021].

\bibitem%[Williams, Turyshev, and Murphy(2005)]
{cospar_llr_04} Williams, J.~G., Turyshev, S.~G., Boggs, D.~H., and Ratcliff, J.~T., ``Lunar Laser Ranging Science: Gravitational Physics and Lunar Interior and Geodesy.'' In proc. ``35th COSPAR Scientific Assembly,'' July 18-24, 2004, Paris, France, 2004. To be published in {\it Adv. Space Res.}, (2005). 

\bibitem%[Williams, Turyshev, and Boggs(2005)]
{Williams_Turyshev_Boggs_2004} Williams, J.~G., Turyshev, S.~G., Boggs, D.~H., ``New Results in LLR Tests o Relativistic Gravity.'' To be published in {\it Phys. Rev. Letters} (2005).
\end{thebibliography}
\end{document}